\documentclass[aps,pra,twocolumn,letterpaper,showpacs,superscriptaddress,floatfix,10pt]{revtex4-1}
\usepackage[pdftex]{graphicx}% Include figure files
\graphicspath{ {./images/} }
\usepackage{dcolumn}% Align table columns on decimal point
\usepackage{bm}% bold math
\usepackage{bbm}% bold math
\usepackage[usenames, dvipsnames]{color}
\usepackage{comment}
\usepackage[colorlinks, linkcolor=blue]{hyperref}

\usepackage{layouts}

\usepackage[T1]{fontenc}

\usepackage{amsfonts}
\usepackage{amssymb}
\usepackage{amsmath}

\usepackage{amsthm}
\theoremstyle{remark}
\newtheorem*{thm}{Theorem}

\usepackage{multirow}
\usepackage[scr=boondoxo, scrscaled=1.05]{mathalfa}

\usepackage{subfigure}
\usepackage{overpic}

\usepackage{array}
\newcolumntype{L}[1]{>{\raggedright\let\newline\\\arraybackslash\hspace{0pt}}m{#1}}
\newcolumntype{C}[1]{>{\centering\let\newline\\\arraybackslash\hspace{0pt}}m{#1}}
\newcolumntype{R}[1]{>{\raggedleft\let\newline\\\arraybackslash\hspace{0pt}}m{#1}}
\usepackage{tabularx}

\def\KeyWord#1{$\backslash$\IfColor{$\!\!$\textRed{#1}\textBlack}{#1}$\!\!$}

\usepackage{wasysym}

\newcommand{\reals}{\mathbb{R}}

\newcommand{\integers}{\mathbb{Z}}

\renewcommand{\i}{\mathrm{i}}
\renewcommand{\d}{\mathrm{d}}

\newcommand{\Bn}{{\vec{n}}}
\newcommand{\Bm}{{\vec{m}}}
\newcommand{\Bk}{{\vec{k}}}

\newcommand{\BO}{{\vec{\Omega}}}
\newcommand{\Bt}{{\vec{\theta}}}

\def\bra#1{\langle#1|}
\def\ket#1{|#1\rangle}

\def\ketbra#1#2{|#1\rangle\langle#2|}

\def\cexp#1{\langle#1\rangle}
\def\ccexp#1{\langle\!\langle#1\rangle\!\rangle}
\def\com#1#2{\left[#1,#2\right]}
\def\tr#1{\mathrm{Tr}\left[#1\right]}

\begin{document}
\title{Nonadiabatic Topological Energy Pumps with Quasiperiodic Driving}

\author{David M. Long}
\email{dmlong@bu.edu}
\affiliation{Department of Physics, Boston University, Boston, Massachusetts 02215, USA}

\author{Philip J. D. Crowley}
\affiliation{Department of Physics, Boston University, Boston, Massachusetts 02215, USA}
\affiliation{Department of Physics, Massachusetts Institute of Technology, Cambridge, Massachusetts 02139, USA}

\author{Anushya Chandran}
\affiliation{Department of Physics, Boston University, Boston, Massachusetts 02215, USA}

\date{\today}

\begin{abstract}
    We derive a topological classification of the steady states of \(d\)-dimensional lattice models driven by \(D\) incommensurate tones.
    Mapping to a unifying \((d+D)\)-dimensional localized model in frequency space reveals anomalous localized topological phases (ALTPs) with no static analog.
    While the formal classification is determined by \(d+D\), the observable signatures of each ALTP depend on the spatial dimension \(d\).
    For each \(d\), with \(d+D=3\), we identify a quantized circulating current, and corresponding topological edge states.
    The edge states for a driven wire (\(d=1\)) function as a quantized, nonadiabatic energy pump between the drives.
    We design concrete models of quasiperiodically driven qubits and wires that achieve ALTPs of several topological classes.
    Our results provide a route to experimentally access higher dimensional ALTPs in driven low-dimensional systems.
\end{abstract}

\maketitle

\emph{Introduction.---} Topological order in static systems can underlie the quantization of nonequilibrium responses in slowly driven systems with fewer dimensions.
A well-known example is the Thouless charge pump, a slowly and periodically driven insulating wire that transmits charge at a quantized rate~\cite{Thouless1983,Switkes1999,Kraus2012,Lohse2016,Nakajima2016,Marra2020}.
Fourier transform of the drive produces a static frequency lattice with one spatial and one synthetic dimension in the presence of a weak electric field.
Charge pumping in the wire can then be understood via the integer quantum Hall effect in the frequency lattice.
Analogous arguments also relate a qubit slowly driven by two incommensurate tones to a frequency lattice with two synthetic dimensions~\cite{Martin2017,Crowley2019,Crowley2020,Boyers2020,Nathan2019c,Nathan2020}.
In this setting, the integer quantum Hall effect manifests as a quantized energy current between the two drives.

Can nonequilibrium responses be quantized away from the adiabatic limit?
An affirmative answer with periodic driving is provided by the two-dimensional anomalous Floquet-Anderson insulator (AFAI), which transports charge along the boundary of the sample at a quantized rate.
Charge pumping in the AFAI can be understood through a topological invariant of a frequency lattice, now with two spatial and one synthetic dimension~\cite{Titum2016,Nathan2017}.

\begin{figure}[b]
    \centering
    \includegraphics[width=\linewidth]{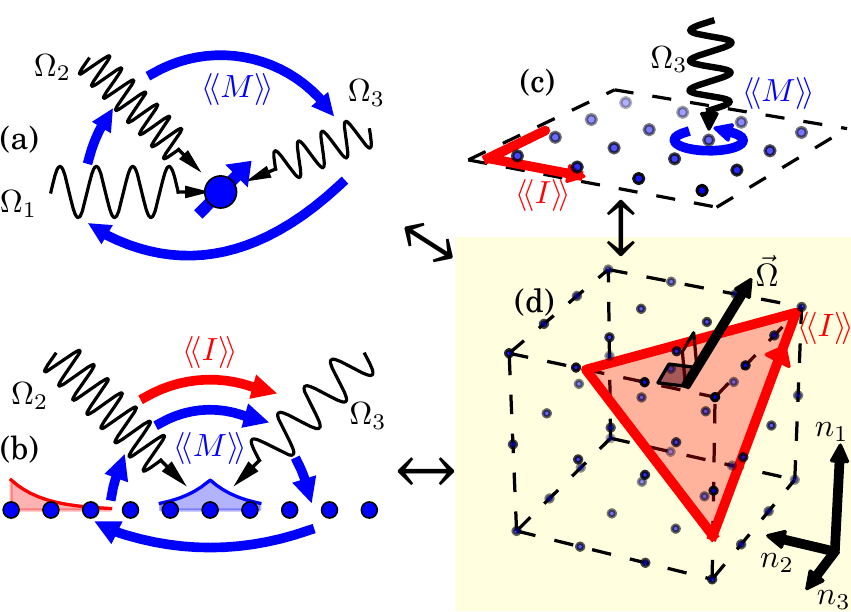}
    \caption{\label{fig:correspond}Correspondence between ALTPs with fixed \(d+D=3\). \textbf{(a)} \((0+3)\): A qudit driven by three incommensurate frequencies showing chiral circulation of energy \(\ccexp{M}\) between the drives. \textbf{(b)} \((1+2)\): A localized fermionic chain driven by two tones also exhibits an energy-charge circulation in the bulk, and topological edge states that pump energy between the drives \(\ccexp{I}\). \textbf{(c)} \((2+1)\): A localized two-dimensional system driven by one tone has a quantized bulk magnetization, and quantized edge currents~\cite{Titum2016,Nathan2017}. \textbf{(d)} \((3+0)\): All three systems have a unifying description in terms of a static frequency lattice with localized bulk eigenstates and an electric field \(\BO\).}
\end{figure}

In this Letter, we use the frequency lattice construction to reveal nonadiabatic topological responses in quasiperiodically driven systems. Our key observation is that the frequency lattice treats spatial and synthetic dimensions on an equal footing when its eigenstates are localized (\autoref{fig:correspond}). More formally, the topological classification of localized phases of \(d\)-dimensional tight-binding models driven by \(D\) incommensurate periodic tones depends only on the total frequency lattice dimension \(d+D\). The classification is by an integer when \(d+D>1\) is odd and is trivial otherwise. We call the nontrivial phases anomalous localized topological phases (ALTPs).

We obtain the observable signatures of ALTPs for each \(d\) with \(d+D=3\) from the frequency lattice as summarized in \autoref{fig:correspond}.
In particular, edge states in the frequency lattice manifest as a quantized energy current between the two drives for the \(d=1\) wire.
We verify our predictions numerically in concrete models, and close with experimental implications.

The effects of quasiperiodic driving have been extensively studied in few-body systems~\cite{Ho1983,Luck1988,Casati1989,Jauslin1991,Blekher1992,Jorba1992,Feudel1995,Bambusi2001,Gentile2003,Chu2004,Gommers2006,Chabe2008,Cubero2018,Nandy2018,Ray2019} and more recently in many-body settings~\cite{Nandy2017,Dumitrescu2018,Kolodrubetz2018,Peng2018,Peng2018b,Zhao2019,Else2019,Friedman2020}. A cohesive frequency lattice lens allows us to significantly expand the number of dynamical phases accessible by quasiperiodic driving.

\emph{Localization and the frequency lattice.---} Tight-binding models in \(d\) spatial dimensions driven by \(D\) incommensurate periodic tones are described by a Hamiltonian \(H(\theta_{d+1}(t),\ldots,\theta_{d+D}(t))\) (we use the indices \(j \in \{1,\ldots,d\}\) for spatial dimensions). We assume \(H\) is smooth and periodic in each of the \(D\) drive phases \(\theta_j(t) = \Omega_j t + \theta_{0j}\), where \(\Omega_{d+i}\) is the angular frequency of the \(i\)th drive and \(\theta_{0(d+i)}\) is its initial phase. For brevity of notation, we assemble the drive phases and frequencies into a vector such as \(\BO = \sum_{j=d+1}^{d+D} \Omega_j \hat{e}_j\). The quasiperiodicity of the driving is stated formally as \(H(\Bt_t) = H(\Bt_t + 2\pi \hat{e}_j)\)~\footnote{Throughout the main text we set \(\Bt_0 = 0\), which is equivalent to shifting the origin of time for observables.}.

We look for a basis of solutions to the Schr\"odinger equation of the form
\begin{equation}
     \ket{\psi_\alpha(t)} = e^{-i\epsilon_\alpha t}\ket{\phi_\alpha(\Bt_t)},
     \label{eqn:quasist_def}
\end{equation}
where \(\alpha\) indexes the system's Hilbert space, \(\epsilon_\alpha\) is a constant quasienergy and \(\ket{\phi_\alpha(\Bt)}\) is a smooth quasienergy state defined on the torus of drive phases. Equation \eqref{eqn:quasist_def} is the generalization of the Floquet-Bloch decomposition to quasiperiodic driving~\cite{Blekher1992,Floquet1883,Ho1983}. Substituting Eq.~\eqref{eqn:quasist_def} into the time-dependent Schr\"odinger equation, \(i\partial_t \ket{\psi_\alpha(t)} = H(\Bt_t) \ket{\psi_\alpha(t)}\) (with \(\hbar =1\)), and Fourier transforming gives
\begin{equation}
    \epsilon_\alpha \ket{\phi_{\alpha\Bn}} = \sum_{\Bm \in \integers^D} \left( H_{\Bn-\Bm} - \Bn \cdot \BO \delta_{\Bn \Bm} \right) \ket{\phi_{\alpha\Bm}},
\end{equation}
where \(H_\Bn\) and \(\ket{\phi_{\alpha\Bn}}\) are the Fourier components of \(H(\Bt)\) and \(\ket{\phi_\alpha(\Bt)}\), respectively.

Introducing auxiliary degrees of freedom associated to the Fourier components \(\ket{\Bn}\) -- a frequency lattice -- allows the quasienergy states to be explicitly represented as the eigenstates \(\sum_{\Bn\in\integers^D}\ket{\phi_{\alpha\Bn}}\ket{\Bn}\) of a quasienergy operator~\cite{Ho1983,Blekher1992,Verdeny2016}:
\begin{equation}
    K = \sum_{\Bn,\Bm \in \integers^D} \left( H_{\Bn-\Bm} - \Bn \cdot \BO \delta_{\Bn \Bm} \right) \ketbra{\Bn}{\Bm}.
    \label{eqn:K}
\end{equation}
\(K\) has the form of a lattice Hamiltonian with an electric field \(\BO\) and translationally invariant hopping matrices \(H_{\Bn}\). As the \(H_{\Bn}\) themselves act on a \(d\)-dimensional lattice, the full dimension of the frequency lattice is \(d+D\), with \(d\) spatial and \(D\) synthetic dimensions. The frequency lattice is illustrated in \autoref{fig:correspond}(d) for \(d+D = 3\).

Although Eq.~\eqref{eqn:K} holds for classical driving, it is useful to interpret the auxiliary state \(\ket{n_j}\) as corresponding to the photon number of the \(j\)th drive~\cite{Nathan2019c}. A nearest-neighbor hop along direction \(j\) then corresponds to photon emission or absorption into drive \(j\), while the potential energy \(\Omega_j n_j\) accounts for the energy of the drive.

We demand localization of the quasienergy states in the spatial dimensions for the purposes of stability of our classification. Localization ensures that small perturbations by local operators do not strongly couple distant quasienergy states, preventing the dramatic rearrangement of eigenstates that could otherwise occur.

Localization in the synthetic dimensions is then equivalent to the existence of a complete set of solutions~\eqref{eqn:quasist_def}. To see this, note that the Fourier expansion of the solution
\begin{equation}
    \ket{\psi_\alpha(t)} = e^{-i\epsilon_\alpha t} \ket{\phi_\alpha(\Bt_t)} = e^{-i\epsilon_\alpha t} \sum_{\Bn\in\integers^D}\ket{\phi_{\alpha\Bn}} e^{-i \Bn \cdot \Bt_t}
    \label{eqn:time_quasist}
\end{equation}
converges only when the Fourier components \(\ket{\phi_{\alpha\Bn}}\) are square summable, that is, when the eigenstates of \(K\) are normalizable. When we have localization in both spatial and synthetic dimensions, we refer to the system as being in a localized phase.

An immediate dynamical consequence of localization is the quasiperiodic time dependence of local observables. This follows from the decomposition of an observable \(O(t)\) in the Heisenberg picture in the basis of quasienergy states,
\begin{equation}
    O(t) = \sum_{\alpha, \beta} 
    O_{\alpha \beta}(\Bt_t) e^{-i(\epsilon_\beta-\epsilon_\alpha)t} 
    \ketbra{\phi_\alpha(\Bt_0)}{\phi_\beta(\Bt_0)},
    \label{eqn:obs_quasi}
\end{equation}
where \(O_{\alpha \beta}(\Bt_t) = \bra{\phi_\alpha(\Bt_t)} O(0) \ket{\phi_\beta(\Bt_t)}\) is quasiperiodic. (We have assumed \(O\) does not have explicit time dependence.)
Roughly, localization of \(\ket{\phi_\alpha(\Bt_t)}\) implies that only finitely many of the terms \(O_{\alpha \beta}(\Bt_t)\) contribute significantly to expectation values, so that \(\cexp{O(t)}\) is explicitly quasiperiodic.
Mathematically, the power spectrum of all local observables in a localized phase is pure point~\cite{Jauslin1991,Blekher1992}.

\begin{figure*}
    \centering
    \includegraphics[width=\linewidth]{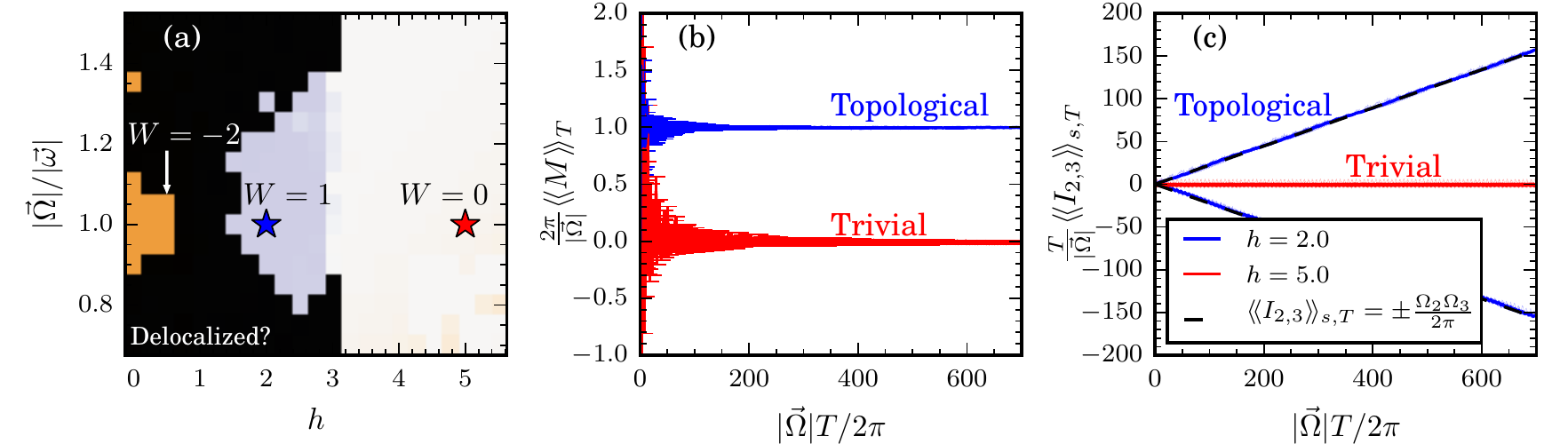}
    \caption{\label{fig:numerics}Numerics for the model \(H^\delta\). \textbf{(a)} \((0+3)\): Phase diagram. Colors represent the value of \(\ccexp{M}_T\) over a fixed time \(T\). In black delocalized regions, the Fourier spectrum of \(\ccexp{M}_T\) is not pure point on numerical timescales. \textbf{(b)} \((0+3)\): \(\ccexp{M}_T\) for parameter values which are marked by stars in \textbf{(a)} against averaging time \(T\). The average circulations converge to the theoretically predicted quantized values. \textbf{(c)} \((1+2)\): The work done on drive \(j\) at one edge of the wire, \(T\ccexp{I_j}_{s,T}\)~\eqref{eqn:en_current}. Energy is transported between drives 2 and 3 at the predicted quantized average rate. Parameters: \textbf{(a)} \(\delta/B_0 = 0.01\), \(\vec{\omega}/B_0 = (2, 1.618\,031..., 1.073\,506...)\) and \(\BO \propto \vec{\omega}\). \textbf{(b)} \(\BO = \vec{\omega}\), \(h = 2\) (blue), and \(h=5\) (red). \textbf{(c)} Lattice size \(L=40\), \(s=14\) sites filled; see Eq. \eqref{eqn:1+2D}. Further details are reported in~\cite{SUPP}.}
\end{figure*}

\emph{Formal classification.---} The topological classification is most naturally expressed through the micromotion operator (cf.~\cite{Roy2017a})
\begin{equation}
    V(\vec{\Phi},\Bt) = \sum_\alpha \ketbra{\phi_\alpha(\Bt)}{\alpha},
\end{equation}
where \(\ket{\alpha}\) is a basis for the system's Hilbert space and we have suppressed the dependence of the quasienergy states on the \(d\) fluxes \(\vec{\Phi}\) twisting the periodic boundary conditions of the spatial dimensions.
In a localized phase the micromotion \(V(\vec{\Phi},\Bt)\) is a smooth map from the \((d+D)\)-dimensional torus defined by \(\vec{\Phi}\) and the drive phases \(\Bt\) to the unitary group. It is well known that such maps are classified by an integer winding number \(W[V]\) when \(d+D\) is odd, defined by~\cite{Nakahara2003,Teo2010,Kitagawa2010,Yao2017}
\begin{equation}
    W[V] = C_{d+D} \int \d^d\Phi\,\d^D\theta\, \epsilon^{j\cdots k} \tr{(V^\dagger \partial_j V)\cdots (V^\dagger \partial_k V)}
\end{equation}
where the integral is over the torus, \(\epsilon^{j\cdots k}\) is the Levi-Civita symbol, \(\partial_j\) is differentiation with respect to one of \(\Phi_j\) or \(\theta_{j}\), and \(C_{d+D}\) is a constant~\cite{SUPP}.

\begin{thm}
    The winding number \(W[V]\) is an integer valued topological invariant characterizing localized phases with \(d+D>1\). That is, if the two Hamiltonian-frequency pairs \((H_0(\Bt),\BO_0)\) and \((H_1(\Bt),\BO_1)\) are joined by a connected path \((H_s(\Bt),\BO_s)\) (where \(s\in [0,1]\)) such that all the \((H_s(\Bt),\BO_s)\) have localized quasienergy states, then \(W[V_0] = W[V_1]\).
\end{thm}

In the Supplemental Material~\cite{SUPP} we show that, under the conditions of the theorem, the path between the micromotion operators \(V_s\) is continuous~\footnote{In the \((0+1)\)-dimensional case of a periodically driven qudit, \(W[V]\) is not a good invariant, as it changes with a gauge transformation of the quasienergy states~\cite{SUPP}. This case is peripheral to our focus on quasiperiodically driven systems.}.
As \(W[V]\) is invariant under smooth deformations of \(V\)~\cite{Nakahara2003}, the theorem follows~\cite{SUPP}. We refer to a localized phase with a nontrivial winding number \(W[V] \neq 0\) as an anomalous localized topological phase (ALTP).

As promised, the classification depends only on the frequency lattice dimension \(d+D\). Note that the Floquet classification of anomalous phases without symmetry is reproduced with \(D=1\)~\cite{Roy2017a}.

\emph{Observable consequences.---} The formal classification of ALTPs is physically interesting only because it predicts quantized observables. The physical observables depend on \(d\). We identify these for \(d \in \{0,1\}\) when \(d+D=3\) and later verify our predictions numerically (\autoref{fig:numerics}). The Floquet case \(d=2\) is well studied~\cite{Titum2016,Nathan2017}.

The chiral energy circulation captures the topological response of \((0+3)\)-dimensional ALTPs -- qudits driven by three incommensurate tones. In more detail, the Heisenberg operator for the instantaneous rate of work done on the qudit is \(U^\dagger\partial_t H(\Bt_t) U = \sum_{j} \Omega_j U^\dagger \partial_{j} H(\Bt_t) U\), where \(U = U(t,0)\) is the evolution operator from time \(0\) to \(t\). As energy input into the qudit must come from the drives, it is natural to identify \(U^\dagger\partial_{j} H U \equiv -\dot{n}_j\) as the rate of photon transfer out of the \(j\)th drive. An operator measuring the rate at which photons circulate between the drives (\autoref{fig:correspond}(a)) is then
\begin{equation}
    M(t) = \frac{1}{4}(\Bn \times \dot{\Bn})\cdot \hat{\Omega} + \mathrm{H.c.},
\end{equation}
where \(\Bn\) is the integral of \(\dot{\Bn}\), and we can drop the constant of integration~\cite{SUPP}. Introducing the notation \(\ccexp{A}_T \equiv \frac{1}{T}\int_0^T \d t\, \tr{A(t)}\), we prove~\cite{SUPP}
\begin{equation}
    \ccexp{M}_T = \frac{|\BO|}{2\pi} W[V] + O(T^{-1}).
    \label{eqn:en_circ}
\end{equation}
That is, the long-time average of the circulation in an initial mixed state \(\rho \propto \mathbbm{1}\) is quantized and proportional to the winding number.

A quantized circulation is also present in \((1+2)\)-dimensional ALTPs -- wires driven by two incommensurate tones. However, in contrast to the \((0+3)\)-dimensional ALTP, there are also \emph{edge} signatures of topology.

At a boundary of the wire there is a topological \emph{energy current} between the drives (\autoref{fig:correspond}(b)). The frequency lattice for the driven wire has a slab geometry. A nonzero winding number in the bulk is accompanied by current-carrying edge states, which must run perpendicular to \(\BO\), due to Stark localization by the electric field \(\BO\) (\autoref{fig:correspond}(d)). If \(\BO = \Omega_2 \hat{e}_2+\Omega_3\hat{e}_3\), the edge current is parallel to \(\Omega_3\hat{e}_2-\Omega_2\hat{e}_3\). That is, photons are transferred from drive 3 to drive 2 (or drive 2 to drive 3, depending on the sign of \(W[V]\)) in an energy current.

Quantitatively, the long-time average of the energy current into drive \(j\), \(I_j(t) = \Omega_j \dot{n}_j(t)\), in an initial state localized near an edge is~\cite{SUPP}
\begin{equation}
    \ccexp{I_j}_{s,T} \equiv \ccexp{I_j \rho_s}_T 
    = \pm \frac{\Omega_2 \Omega_3}{2\pi} W[V] + O(T^{-1},e^{-s/\xi}).
    \label{eqn:en_current}
\end{equation}
Here, \(\rho_s\) is a projector onto lattice sites localized within \(s\) sites of the edge, \(\xi\) is the single-particle localization length, and the sign depends on which drive \(j\) is being considered. Experimentally, this is the response of a noninteracting wire filled with fermions up to a distance \(s\) from the edge.

\emph{Model.---} Constructing a \((0+3)\)-dimensional ALTP is difficult because simultaneously achieving localization in all the (synthetic) frequency lattice dimensions and a nonzero winding number is delicate. The electric field \(\BO\) must be the cause of localization, as in the absence of \(\BO\) the quasienergy operator \(K\) is translationally invariant and all eigenstates are delocalized Bloch states. Indeed, a strong \(\BO\) compared to the typical hopping amplitude \(J\) does localize the quasienergy states, as adjacent sites in the frequency lattice become far detuned from one another. However, too short a localization length cannot lead to a nonzero circulation~\eqref{eqn:en_circ}. By adding further-neighbor hops, we engineer a ``sweet spot'' with localization across a few sites and quantized circulation.

We construct a family of driven qubit models indexed by \(\delta\geq 0\). We define \(H^{\delta=0}(\Bt) = -(\vec{B}_1+\vec{B}_{2})\cdot\vec{\sigma}/2\), where \(\vec{\sigma}\) is the vector of Pauli matrices, \(\vec{B}_1\) corresponds to a short-range model on the frequency lattice and \(\vec{B}_2\) has the further-neighbor hops. Explicitly, \(\vec{B}_1 = B_0 \bra{\eta} \vec{\sigma} \ket{\eta}\) where
\begin{multline}
    \ket{\eta(\Bt)} = [\sin\theta_1 + i\sin\theta_2]\ket{\!\uparrow} \\
    + \left[\sin\theta_3 + i(h + \sum_{k=1}^3 \cos\theta_k) \right]\ket{\!\downarrow}
\end{multline}
is an unnormalized eigenstate of \(\vec{B}_1 \cdot \vec{\sigma}\), and \(h\) is a dimensionless parameter~\cite{Deng2013,Unal2019}. The field \(\vec{B}_2\) is given by
\begin{equation}
    \vec{B}_{2} = \frac{[(\vec{\omega} \cdot \nabla)\vec{B}_1]\times\vec{B}_1}{|\vec{B}_1|^2}.
\end{equation}
When the vector of parameters \(\vec{\omega} = \BO\), this term acts as a counterdiabatic correction ensuring that the quasienergy states are parallel to \(\ket{\eta(\Bt)}\)~\cite{Sels2017,Guery2019}. Our choice of \(\ket{\eta(\Bt)}\) then allows for nonzero \(W[V]\), depending on the value of \(h\)~\cite{Deng2013,Unal2019}. For general \(\delta >0\) we truncate the Fourier spectrum of \(H^{\delta=0}(\Bt)\):
\begin{equation}
    H^{\delta}(\vec{\theta}) = \sum_{\{\Bn \,:\, \|H_{\Bn}^0\|_F \geq \delta\}} e^{-i \Bn\cdot\vec{\theta}} H_{\Bn}^0,
    \label{eqn:Hdel}
\end{equation}
where \(H_{\Bn}^0\) are the Fourier coefficients of \(H^{\delta=0}(\Bt)\) and \(\|\cdot\|_F\) is the Frobenius norm.

Figure \ref{fig:numerics}(a) shows the phase diagram obtained by numerically solving the Schr\"odinger equation for \(H^{\delta>0}\) with fifth-nearest-neighbor hops~\footnote{The \(H^{\delta}\) used in our numerics has exponentially decaying hops up to fifth-nearest neighbor in the maximum norm, or eighth-nearest-neighbor in the 1-norm.}. Three topological phases of winding numbers \(W \in \{0,1,-2\}\) are visible. The quantized energy circulation of these phases~\eqref{eqn:en_circ} is verified in \autoref{fig:numerics}(b).

The winding numbers coincide with those predicted by having quasienergy states parallel to \(\ket{\eta(\Bt)}\) near \(\BO=\vec{\omega}\). For large enough frequency \(|\BO| \gg B_0,|\vec{\omega}|\) the localization length is short and \(W = 0\) (not shown). At lower frequencies \(|\BO|\lesssim B_0,|\vec{\omega}|\) the quasienergy states may delocalize.

\begin{figure}
    \centering
    \includegraphics[width=\linewidth]{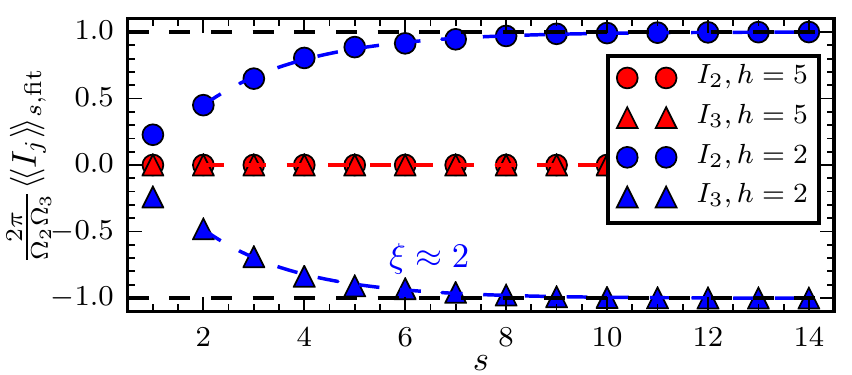}
    \caption{\label{fig:edge_loc}Localization of edge modes. A fit to the slope of \(T\ccexp{I_j}_{s,T}\) with \(T\) gives an estimate of the average current, \(\ccexp{I_j}_{s,\mathrm{fit}}\). The current increases exponentially toward the predicted value~\eqref{eqn:en_current} as the number of initially filled sites \(s\) is increased. An exponential fit in the topological regime gives a localization length of \(\xi \approx 2\). Parameters: as in \autoref{fig:numerics}(c). Further details are reported in~\cite{SUPP}.}
\end{figure}

By reinterpreting one of the synthetic dimensions of \(H^\delta(\Bt)\) as spatial we obtain a \((1+2)\)-dimensional model with ALTPs. Then the electric field causes localization in the spatial dimension. Explicitly, drive 1 (say) is replaced with a lattice of sites \(\ket{n_1}\) with an electric potential \(-\Omega_1 n_1\) and quasiperiodically time-dependent \(m\)-site hops: 
\begin{equation}
    H^\delta_{m}(\theta_2, \theta_3) = \int \frac{\d \theta_1}{2\pi} e^{i m \theta_1} H^\delta(\Bt).
    \label{eqn:1+2D}
\end{equation}
Truncating the lattice provides the necessary edges to observe topological boundary effects.

Figure \ref{fig:numerics}(c) confirms that the energy current is quantized~\eqref{eqn:en_current}. The exponential localization of the current-carrying modes at the edge can be observed in \autoref{fig:edge_loc}.

\emph{Experimental prospects.---} \((0+D)\)-dimensional ALTPs -- driven qudits -- are within immediate experimental reach in a number of solid-state and optical architectures~\cite{Kjaergaard2020,Leibfried2003,Doherty2013}.
Signatures of topology in the adiabatic limit have already been observed with two-tone driven nitrogen vacancy centers~\cite{Boyers2020}.

\((1+2)\)-dimensional ALTPs could be achieved in driven fermionic wires~\cite{Nakajima2016}, but equivalent single-particle physics are available in several platforms~\cite{Bloch2008,Duan2010,Kjaergaard2020,Roushan2017,Ozawa2019b,Maynard2001}, in particular, a qubit coupled to a quantum cavity~\cite{Ozawa2019a} or a bosonic chain. In the former the ``edge'' at which an energy current occurs is the vacuum state. We demonstrate topological signatures in the qubit-cavity ALTP in~\cite{SUPP}.

An energy current between two drives can form the basis of many useful devices, e.g., conversion of photons from one frequency to another~\cite{Martin2017,Peng2018b,Kolodrubetz2018,Crowley2019,Crowley2020}, cooling cavities~\cite{Nathan2019c}, one-way circulation of energy between cavities and the preparation of exotic cavity states~\cite{Nathan2019c,LongTBD}.
The nonadiabatic energy currents of the \((1+2)\)-ALTPs promise to make these devices faster and more stable.

\emph{Outlook.---} We have derived a stable topological classification of quasiperiodically driven lattice models and identified the quantized observables that distinguish each phase. Localization in a dual frequency lattice is central to our understanding of these phases.

In one dimension, localization is believed to be stable to the addition of weak interactions~\cite{Anderson1958,Basko2006,Oganesyan2007,Abanin2019}.
We speculate that \(d=1\) ALTPs are stable interacting phases, even at infinite temperature, due to this many-body localization (MBL); see Refs.~\cite{Nathan2019a,Nathan2019b} for the \((2+1)\) case.
The MBL ALTPs would provide new examples of localization-protected quantum order~\cite{Huse2013}.
The bulk l bits of the MBL chain would support an energy-charge circulation, while the edge l bits would support energy currents.

We have discussed observable signatures for each ALTP with \(d+D = 3\). A natural extension is determining the corresponding observables for larger numbers of drives and other symmetry classes. This would provide experimental access to the topological physics of driven systems in four dimensions and higher. To date, such responses have been observed only in the adiabatic limit~\cite{Lohse2018,Zilberberg2018}.

The authors are grateful to I. Martin and C. Laumann for several helpful discussions. We would also like to thank E. Boyers, D. Else, M. Kolodrubetz, Y. Peng, M. Rudner and A. Sushkov. Numerics were performed on the BU Shared Computing Cluster. This research was supported by NSF Grant No. DMR-1752759 and AFOSR Grant No. FA9550-20-1-0235.

\emph{Note added.---} Recently, Ref.~\cite{Nathan2020b} appeared, which provides complementary models of \((1+2)\)- and \((0+3)\)-dimensional ALTPs. Where Ref.~\cite{Nathan2020b} overlaps with this work, they agree. 

\bibliography{nonad_top_pump_qp}

\clearpage

\section*{Supplemental Material}
    \label{sec:supp}

\subsection{The Frequency Lattice}
    \label{sec:freq_latt}

    In this section we review the frequency lattice construction in detail. This provides a formal mapping between a tight-binding model in \(d\) dimensions driven by \(D\) incommensurate tones and a static system with an additional \(D\) synthetic dimensions. We first review the construction for the case of a single periodic drive (\(D = 1\)) and recover familiar results of Floquet theory. We then make a straightforward generalization to the multiple-tone case.

    We are concerned with the dynamics of driven models in \(d\) dimensions. The state vector \(\ket{\psi(t)}\) of this system obeys the Schr\"odinger equation (with \(\hbar = 1\))
    \begin{equation}
        i\partial_t \ket{\psi(t)} = H(t)\ket{\psi(t)},
        \label{eqn:Schordinger}
    \end{equation}
    where \(H(t)\) is the time dependent Hamiltonian.

    It is frequently useful to describe evolution in terms of a unitary evolution operator \(U(t,t')\) such that \(U(t_1,t_0)\ket{\psi(t_0)} = \ket{\psi(t_1)}\). This operator also obeys the Schr\"odinger equation \(i\partial_{t_1} U(t_1,t_1) = H(t_1)U(t_1,t_0)\), and is often expressed as the time ordered exponential
    \begin{equation}
        U(t_1,t_0) = \mathcal{T}\exp\left(-i\int_{t_0}^{t_1} H(s) \, \d s\right).
    \end{equation}

    \subsubsection{One Drive -- Floquet Theory}

        When the driving is periodic, \(H(t+T) = H(t)\), it is possible to say more about the structure of \(U(t_1,t_0)\). In this case it is possible to identify a complete set of solutions to the Schr\"odinger equation of the form
        \begin{equation}
            \ket{\psi_\alpha (t)} = e^{-i\epsilon_\alpha t}\ket{\phi_\alpha(t)}
            \label{eqn:qe_soln}
        \end{equation}
        where \(\ket{\phi_\alpha(t+T)} = \ket{\phi_\alpha(t)}\) is periodic and \(\alpha\) indexes a basis of the Hilbert space. Due to the similarity in form to the evolution of an eigenstate of a static Hamiltonian, \(\epsilon_\alpha\) is called the quasienergy and \(\ket{\phi_\alpha(t)}\) is called the quasienergy state.

        The decomposition of the solutions~\eqref{eqn:qe_soln} implies the corresponding decomposition of the evolution operator
        \begin{equation}
            U(t_1,t_0) = V(t_1)e^{-i (t_1-t_0)H_F }V^\dagger(t_0),
            \label{eqn:FBD}
        \end{equation}
        where \(H_F = \sum_{\alpha} \epsilon_\alpha \ketbra{\alpha}{\alpha}\) is called the Floquet Hamiltonian, the micromotion \(V(t) = \sum_{\alpha} \ketbra{\phi_\alpha(t)}{\alpha}\) is also periodic and \(\ket{\alpha}\) is an arbitrary fixed basis for the system's Hilbert space. This decomposition is the subject of \emph{Floquet's theorem}~\cite{Floquet1883}, and we will call it a \emph{Floquet decomposition}. We will prove the same result using the frequency lattice.

        Before we do so, it is important to note that there is a gauge freedom in this decomposition; as \(e^{in \Omega t}\) for \(n\in\integers\) and \(\Omega = 2\pi/T\) is itself periodic with period \(T\), the form of the Floquet decomposition is preserved by the map
        \begin{equation}
            \epsilon_\alpha \mapsto \epsilon_\alpha + n_\alpha \Omega, \quad
            \ket{\phi_\alpha(t)} \mapsto e^{i n_\alpha \Omega t}\ket{\phi_\alpha(t)}.
            \label{eqn:floq_gauge}
        \end{equation}
        The quasienergy can be shifted by an integer multiple of \(\Omega\) without affecting the actual solution~\eqref{eqn:qe_soln}, provided a gauge transformation is made to the quasienergy state \(\ket{\phi_\alpha(t)}\). As such, the quasienergy should be regarded as being defined modulo \(\Omega\). This is the origin of the distinct topology possible in the structure of Floquet systems when compared to static systems~\cite{Rudner2013,Nathan2015,Roy2017a}.

        Floquet's theorem can be proved using elementary techniques in linear ordinary differential equations, but these techniques are not easily transferable to the quasiperiodic case. We will now describe how the time dependent Floquet problem can be understood as a static lattice problem with one synthetic dimension -- the frequency lattice -- and how this interpretation naturally leads to Floquet's theorem.

\begin{figure}
            \centering
            \includegraphics{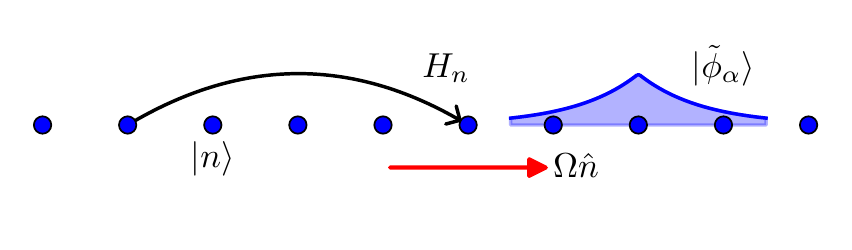}
            \caption{\label{fig:frqlat_floq}\emph{Frequency lattice for periodic driving.---} The problem of finding quasienergy states in a periodically driven (Floquet) system can be mapped onto a static frequency lattice problem with one synthetic dimension, with sites labeled by \(\ket{n}\). The lattice Hamiltonian -- the quasienergy operator \(K(\theta_0)\) -- has a linear potential \(-n \Omega\), as might arise from an electric field \(\Omega \hat{n}\), and hopping matrices given by the Fourier components \(H_m\) of \(H(\theta)\). Floquet's theorem follows from the localization of the frequency lattice eigenstates \(\ket{\tilde{\phi}_\alpha}\), which can be seen as a consequence of Stark localization by \(\Omega \hat{n}\).}
        \end{figure}

        By Fourier transforming the Schr\"odinger equation we can map~\eqref{eqn:Schordinger} into a lattice problem in frequency space. That is, if we write \(H(t) = H(\theta_t)\) with \(\theta_t = \Omega t + \theta_0\) defined modulo \(2\pi\), then we can express \(H\) in terms of its Fourier series as
        \begin{equation}
            H(\theta) = \sum_{m\in\integers} H_m e^{-i m \theta}.
        \end{equation}
        Similarly writing \(\ket{\phi_\alpha(\theta_t)} = \sum_{m\in \integers} \ket{\phi_{\alpha m}(\theta_0)} e^{-i m \theta_t}\), the Schr\"odinger equation becomes
        \begin{equation}
            \epsilon_\alpha \ket{\phi_{\alpha n}(\theta_0)} = 
            \sum_{m\in\integers} \left( H_{n-m} e^{-i(n-m)\theta_0} - n\Omega \delta_{nm} \right)
            \ket{\phi_{\alpha m}(\theta_0)}.
            \label{eqn:quasi_schro}
        \end{equation}
        This is the form of a tight-binding model on a one-dimensional lattice with sites labeled by \(n\), and a local Hilbert space given by that of the original system. Indeed, defining an auxiliary Hilbert space spanned by \(\ket{n}\), and defining
        \begin{align}
            \ket{\tilde{\phi}_\alpha(\theta_0)} &= \sum_{n\in\integers} \ket{\phi_{\alpha m}(\theta_0)}\otimes\ket{n}, \\
            K(\theta_0) &= \sum_{n,m\in\integers} \left(H_{n-m} e^{-i(n-m)\theta_0} - n\Omega \delta_{nm} \right) \otimes\ketbra{n}{m},
            \label{eqn:floq_quasien_op}
        \end{align}
        then~\eqref{eqn:quasi_schro} becomes an eigenvalue equation for the \emph{quasienergy operator} \(K(\theta_0)\), which has the form of a lattice Hamiltonian (\autoref{fig:frqlat_floq}).

        We have kept the dependence of \(\ket{\phi_{\alpha n}(\theta_0)}\) and \(K(\theta_0)\) on the initial phase (equivalently initial time) explicit, though we have not written it for \(\epsilon_\alpha\). Indeed, inspecting~\eqref{eqn:floq_quasien_op} shows that the initial phase \(\theta_0\) enters the quasienergy operator like a constant vector potential. This is a pure gauge term, unless our synthetic lattice has non-contractible loops, which it does not. Thus the quasienergies \(\epsilon_\alpha\) can't depend on initial phase. The dependence of the quasienergy states \(\ket{\tilde{\phi}_\alpha(\theta_0)}\) on \(\theta_0\) only encodes the choice of the origin of time in the temporal domain.

        The other terms in \(K(\theta_0)\) also admit translation into the language of a tight-binding model. The Fourier amplitudes \(H_m\) are \(m\)th nearest-neighbor hopping terms, and when \(H(t)\) is smooth in the sense that its Fourier components \(H_m \lesssim e^{-\kappa m}\) (\(\kappa\) being a positive dimensionless constant) decay exponentially, then the lattice model defined by \(K\) is also quasilocal. The term \(\sum_{n} -n\Omega \ketbra{n}{n}\) is an on-site potential of constant gradient \(-\Omega \hat{n}\). This gives \(\Omega\hat{n}\) a natural interpretation as an electric field (in units where \(\hbar = e = 1\)). This geometry is shown in \autoref{fig:frqlat_floq}.

        The presence of the electric field \(\Omega\hat{n}\) implies that the eigenstates of \(K\) must be exponentially localized by Stark localization. Thus, their Fourier transforms \(\ket{\phi_\alpha(\theta)}\) are well defined and smooth as functions of \(\theta\). This is the proof of the existence of the quasienergy states via the frequency lattice -- that is, Floquet's theorem.

        While this is a less elementary than possible proof of Floquet's theorem, the frequency lattice construction has been used in Floquet theory many times before. For instance, various high-frequency expansions for the Floquet Hamiltonian and quasienergy states can be obtained through conventional perturbation theory in the frequency lattice~\cite{Eckardt2015}.

        One may be concerned that we have introduced many more degrees of freedom in our problem than are physical. For example, for a driven qudit with only \(N\) states, \(K\) certainly has many more eigenstates than \(N\). These extra quasienergy states are a result of the gauge invariance~\eqref{eqn:floq_gauge}; the eigenstates of \(K\) fall into \(N\) classes related by translation in the frequency lattice (multiplication by \(e^{in\Omega t}\) in the time domain) and a shift in quasienergy due to the change in potential from the electric field \(\Omega \hat{n}\). This is exactly the transformation~\eqref{eqn:floq_gauge}. In the frequency lattice language, the eigenstates of \(K\) form a Stark ladder, and the gauge freedom relates states on different rungs of the ladder.

        We will see in the next section of this supplementary material that the frequency lattice picture can be adapted to the case of driving by multiple tones. This picture will allow us to import our intuition and known results about static lattice problems to quasiperiodically driven tight-binding models.

    \subsubsection{The Frequency Lattice For Multiple Tones}
        \label{subsec:freq_lat_multi}

        In the multi-tone case, the Hamiltonian is not necessarily periodic, but has the structure
        \begin{equation}
            H(t) = H(\Bt_t) = H(\BO t + \Bt_0)
        \end{equation}
        where the phase angles \(\Bt_t\) should be considered to be defined on a torus; \(\Bt \in \mathbb{T}^D = \reals^D/2\pi\integers^D\).

        When the frequencies \(\BO\) are incommensurate in the sense that \(\Bn\cdot\BO = 0\) only when \(\Bn = 0\) for \(\Bn\in\integers^D\), then Floquet's theorem does not apply. However, we can still make use of the frequency lattice to understand the structure of the solutions to the Schr\"odinger equation.

        To find a Floquet decomposition in analogy to~\eqref{eqn:qe_soln}, we are seeking solutions to the Schr\"odinger equation of the form
        \begin{equation}
            \ket{\psi_\alpha(t;\Bt_0)} = e^{-i\epsilon_\alpha(\Bt_0)}\ket{\phi_\alpha(\Bt_t)},
            \label{eqn:qe_soln_quasi}
        \end{equation}
        where we have kept the dependence \(\epsilon_\alpha(\Bt_0)\) for now in order to treat commensurate and incommensurate drives within the same formalism. Fourier transforming the Schr\"odinger equation with respect to time gives a \(D\)-dimensional lattice model
        \begin{equation}
            \epsilon_\alpha(\Bt_0)\ket{\tilde{\phi}_\alpha(\Bt_0)} = K(\Bt_0)\ket{\tilde{\phi}_\alpha(\Bt_0)}
        \end{equation}
        where
        \begin{align}
            &\ket{\tilde{\phi}_\alpha(\Bt_0)} = \sum_{\Bn} \ket{\phi_{\alpha \Bn}(\Bt_0)} \otimes \ket{\Bn}, \\
            &K(\Bt_0) = \sum_{\Bn,\Bm} \left(H_{\Bn-\Bm} e^{-i(\Bn-\Bm)\cdot\Bt_0} - \Bn\cdot\BO \delta_{\Bn\Bm} \right) \otimes\ketbra{\Bn}{\Bm},
            \label{eqn:multi_quasien_op}
        \end{align}
        and \(H_{\Bn}\) (\(\ket{\phi_{\alpha \Bn}(\Bt_0)}\)) labels the Fourier components of \(H(\Bt)\) (\(\ket{\phi_\alpha(\Bt)}\)) as before.

        If the frequencies are incommensurate, then the frequency lattice again has no non-trivial loops, and the vector potential \(\Bt_0\) is again purely a gauge choice which does not affect the spectrum of \(K(\Bt_0)\). Once more, the only effect of initial phase \(\Bt_0\) on \(\ket{\tilde{\phi}_\alpha(\Bt_0)}\) in the incommensurate case is to encode the origin of time.

        \begin{figure}
            \centering
            \includegraphics{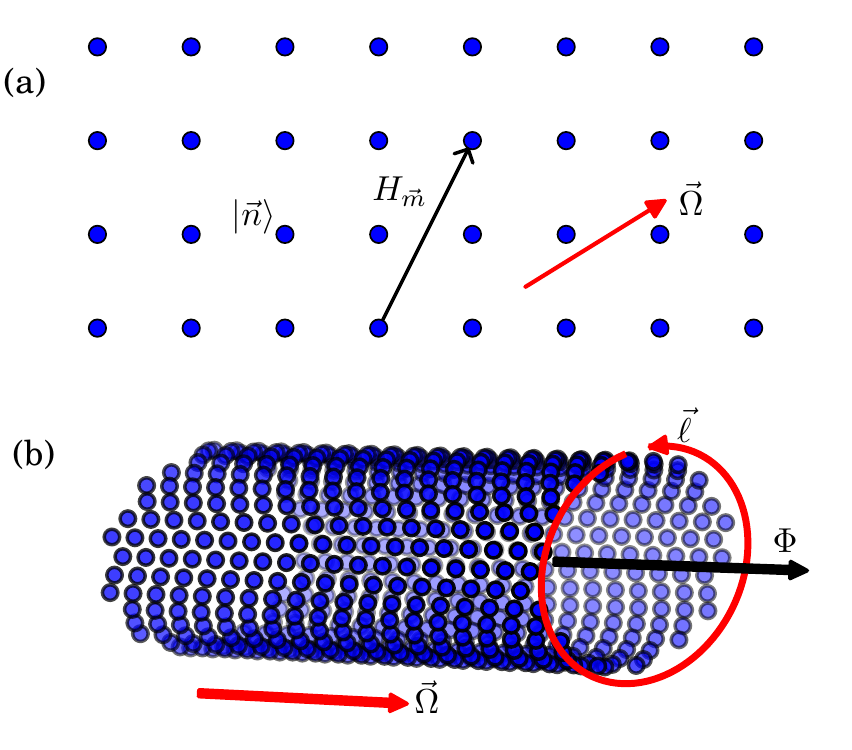}
            \caption{\label{fig:frqlat_quasi}\emph{Frequency lattice for \(D\) tones.---} \textbf{(a)} The quasienergy states of~\eqref{eqn:qe_soln_quasi} can be identified from the eigenstates of a quasienergy operator \(K\)~\eqref{eqn:multi_quasien_op} in a frequency lattice with \(D\) synthetic dimensions (illustrated for \(D=2\)), with sites labeled by \(\ket{\Bn}\). \(K\) consists of hopping matrices \(H_{\Bm}\) between sites separated by \(\Bm\) and an electric field \(\BO\). \textbf{(b)} When \(\BO\) is commensurate, so that \(\vec{\ell}_j \cdot \BO = 0\) for some integer \(\vec{\ell}_j \neq 0\), the frequency lattice compactifies into a cylinder with circumference \(\vec{\ell}_j\). In addition to the hopping matrices and electric field of the incommensurate case, there is now a flux \(\Phi_j = \vec{\ell}_j \cdot \Bt_0\) proportional to the initial phase of the drives which threads the cylinder.}
        \end{figure}

        The rest of the structure of \(K\) is essentially the same as the Floquet case. The Fourier amplitudes \(H_{\Bm}\) appear as hopping amplitudes along the vector \(\Bm\), and the smoothness of \(H(\Bt)\) translates into the hopping being quasilocal. The driving frequencies \(\BO\) appear as an electric field, and thus imply that the quasienergy states \(\ket{\tilde{\phi}_\alpha(\Bt_0)}\) are Stark localized along the \(\BO\) direction whenever \(|\BO|>0\) (\autoref{fig:frqlat_quasi}(a)). Unlike the periodic case, this Stark localization is not sufficient to conclude that the time domain states \(\ket{\phi_\alpha(\Bt_t)}\) are well defined. The frequency lattice states may be delocalized along other directions in the frequency lattice, preventing their Fourier series from converging to a continuous function. At a technical level, it is the convergence or non-convergence of this Fourier series which determines whether we can find a Floquet decomposition~\eqref{eqn:qe_soln_quasi} for the quasiperiodically driven system.

        A similar gauge freedom to the periodic case appears here, as it must for this prescription to make any sense; otherwise we would have many more solutions to the Schr\"odinger equation than there are states in the system's Hilbert space. Just as in the periodic case, any eigenstates of \(K\) may be translated by \(-\Bn\) to obtain a new eigenstate of \(K\) with an energy shifted by \(\Bn\cdot\BO\) due to the change of its position in the electric potential. In the time domain, this corresponds to the transformation
        \begin{equation}
            \epsilon_\alpha \mapsto \epsilon_\alpha + \Bn\cdot\BO, \quad
            \ket{\phi_\alpha(\Bt)} \mapsto e^{i \Bn\cdot\Bt} \ket{\phi_\alpha(\Bt)},
            \label{eqn:quasi_gauge}
        \end{equation}
        which only introduces a static phase to the actual solution \(\ket{\psi_\alpha(t)}\). Thus, the quasienergy should be regarded as being defined only modulo \(\integers^D \cdot \BO\). In the quasiperiodic case the set \(\integers^D \cdot \BO\) is dense in the real line, making the actual value of the quasienergy rarely useful, and unobservable in any experiment.

        The frequency lattice helps us to unpack and separate this gauge freedom. In later sections we will use \(\alpha\) to index the driven system's Hilbert space and \(\gamma\) to index the expanded frequency lattice Hilbert space. Equivalently, \(\alpha\) may index equivalence classes of \(\ket{\tilde{\phi}_\gamma}\) under the gauge transformation~\eqref{eqn:quasi_gauge}. We must keep in mind that most of the degrees of freedom \(\gamma\) in the lattice problem are only calculational tools, and shouldn't affect any prediction we make about the time domain.

        There is a subtlety in case of \emph{commensurate} drives that is absent in~\eqref{eqn:floq_quasien_op}. If the frequencies \(\BO\) are rationally dependent (commensurate), so that there is some \(\vec{\ell}\in\integers^D\) such that \(\vec{\ell}\cdot\BO = 0\), then the time-domain frequencies \(\Bn\cdot\BO\) and \((\Bn+\vec{\ell})\cdot\BO\) are the same, and should not be regarded as distinct in the frequency lattice. Thus, rational dependencies \(\vec{\ell}_j \neq 0\) between the frequencies \(\BO\) compactify the full frequency lattice from \(\integers^D\) to a cylinder with closed non-contractible loops given by \(\vec{\ell}_j\) (\autoref{fig:frqlat_quasi}(b)).

        In this case one must keep the explicit dependence of \(\epsilon_\alpha(\Bt_0)\) on initial phase. As in the periodic case, the initial phase appears in \(K(\Bt_0)\) as a constant vector potential. However, due to the presence of these non-contractible loops in the frequency lattice, this \emph{can} affect the spectrum through the presence of the physically measurable fluxes
        \begin{equation}
            \Phi_j = \vec{\ell}_j \cdot \Bt_0.
        \end{equation}
        It would then be more correct to write \(\epsilon_\alpha(\Bt_0) = \epsilon(\vec{\Phi})\). In particular, \(\epsilon_\alpha(\Bt_0) = \epsilon_\alpha(\BO t + \Bt_0)\) is still constant along trajectories in the phase torus.

        The quasienergy then forms \emph{bands}, again in complete analogy to non-interacting solid-state systems. The fluxes \(\Phi_j\) parameterize some dependence of the quasienergies \(\epsilon_\alpha(\vec{\Phi})\) on the initial relative phase. Many of the usual ideas of band theory will find application in this setting, but will not be relevant to the incommensurate driving we focus on in this work.

\subsection{Topological classification of localized phases}
    \label{sec:top_inv}

    In this section we prove the theorem stated in the main text, and provide further details about the topological invariant classifying anomalous localized topological phases (ALTPs).

    In the main text we claimed that the topological properties of localized phases of quasiperiodically driven tight-binding models are captured by the topology of the \emph{micromotion operator}
    \begin{equation}
        V(\Bt) = \sum_{\alpha} \ketbra{\phi_\alpha(\Bt)}{\alpha},
        \label{eqn:micro}
    \end{equation}
    where \(\ket{\alpha}\) is a fixed reference basis for the system's Hilbert space.
    The micromotion is regarded as a map from the \(d+D\) dimensional torus parameterized by the \(d\) fluxes twisting the periodic boundary conditions of the spatial dimensions, \(\Phi_j\), and the \(D\) drive phases, \(\theta_{j}\), to the unitary group. We assemble all of these into a single vector
    \begin{equation}
        \Bt = \sum_{j=1}^d \Phi_j \hat{e}_j + \sum_{j=d+1}^{d+D} \theta_{j} \hat{e}_j.
        \label{eqn:phi_n_tht}
    \end{equation}
    In the spirit of having a cohesive language for spatial and synthetic dimensions, we are overloading the notation \(\Bt\) with the fluxes from the spatial dimensions. All our previous formulae are consistent with this notation.

    It is well known that the (stable) homotopy class of such maps is characterized by the integer~\cite{Nakahara2003,Teo2010,Kitagawa2010,Yao2017}
    \begin{equation}
        W[V] = C_{d+D} \int_{\mathbb{T}^{d+D}} \d^{d+D}\theta \epsilon^{j\cdots k}
        \tr{(V^\dagger \partial_j V) \cdots (V^\dagger \partial_k V)},
        \label{eqn:wind_num}
    \end{equation}
    where the integral is over the torus, \(\epsilon^{j\cdots k}\) is the Levi-Civita symbol, \(\partial_j\) is differentiation with respect to one of \(\Phi_j\) or \(\theta_{j}\) and
    \begin{equation}
        C_{d+D} = \frac{(\tfrac{d+D-1}{2})!}{(d+D)! (2\pi i)^{(d+D+1)/2}}
    \end{equation}
    is a constant.

    Strictly speaking, this characterizes maps from the \((d+D)\)-sphere to the unitary group. We are ignoring lower homotopy groups of the unitary group and focusing on this so called \emph{strong} invariant~\cite{Kitaev2009b,Teo2010,Roy2017a}.

    \subsubsection{Gauge invariance of \(W[V]\)}
        \label{subsec:gauge}

        The micromotion operator~\eqref{eqn:micro} is not unique. It changes under the gauge transformation of the quasienergy states. If \(W[V]\) is to have any physical meaning, it cannot change under the gauge transformation~\eqref{eqn:quasi_gauge}. We prove this is so below for \(d+D > 1\).

        In the \(d+D = 1\) case, \(W[V]\) gives the familiar winding number \(\tfrac{1}{2\pi i}\oint \tfrac{\d z}{z}\) for the complex number \(z = \det(V)\). This can be altered by an arbitrary integer through the gauge transformation~\eqref{eqn:quasi_gauge}, and so \(W[V]\) has no physical meaning for a \((0+1)\)-dimensional localized phase -- a periodically driven qudit. Even so, an integer classification of zero-dimensional Floquet systems has been reported in, for instance,~\cite{Roy2017a}.

        \(W[V]\) is always zero for \(d+D\) even, so we will focus on the non-trivial case of \(d+D\) being odd.

        The gauge invariant unitary operator characterizing the system is the evolution operator:
        \begin{align}
            U(t_1,t_0; \Bt_0) &= \sum_{\alpha} \ketbra{\psi_\alpha(t_1;\Bt_0)}{\psi_\alpha(t_0;\Bt_0)} \\
            &= V(\Bt_{t_1}) e^{-i(t_1-t_0)H_F} V^\dagger(\Bt_{t_0}),
            \label{eqn:evo}
        \end{align}
        where \(H_F = \sum_{\alpha} \epsilon_\alpha \ketbra{\alpha}{\alpha}\) is the Floquet Hamiltonian. Any transformation of \(V\) and \(H_F\) which preserves the form of this decomposition does not affect the physical operator \(U\).

        Transformations preserving~\eqref{eqn:evo} include the gauge transformations~\eqref{eqn:quasi_gauge} and rotations of the reference basis \(\ket{\alpha} \mapsto \ket{\beta_\alpha}\). We can handle both of these operations at once by writing
        \begin{equation}
            V(\Bt) \mapsto V(\Bt) \tilde{U}(\Bt)
            \label{eqn:V_gauge}
        \end{equation}
        where \(\tilde{U}(\Bt) = \sum_{\alpha} e^{i \Bn_\alpha \cdot \Bt}\ketbra{\alpha}{\beta_\alpha}\) is unitary and \(\Bn_\alpha\) are arbitrary vectors of integers.

        It is convenient to express \(W[V]\) in a coordinate independent form. In the language of differential forms, if we define \(\tilde{n}_V = -i V^\dagger \d V\), where \(\d\) is the exterior derivative, then \(W[V]\) is expressed
        \begin{align}
            W[V] &= i^{d+D}C_{d+D} \int_{\mathbb{T}^{d+D}} \tr{\tilde{n}_V^{\wedge (d+D)}} \\
            &= \tilde{C}_{d+D} \int_{\mathbb{T}^{d+D}} \tr{\tilde{n}_V \wedge (i\d\tilde{n}_V)^{\wedge (d+D-1)/2}}, \label{eqn:wind2}
        \end{align}
        where \(\tilde{C}_\delta = i^\delta C_\delta\) and we used the fact that
        \begin{multline}
            \d \tilde{n}_V = -i \mathbbm{1} \d (V^\dagger \d V) = -i V^\dagger V \d V^\dagger \wedge \d V \\
            = i V^\dagger \d V \wedge V^\dagger \d V = -i \tilde{n}_V \wedge \tilde{n}_V.
        \end{multline}
        The second equality follows from \(V^\dagger V = \mathbbm{1}\) and \(\d(V^\dagger \d V) = \d V^\dagger \wedge \d V + V^\dagger \d^2 V\) with \(\d^2 = 0\). The third equality uses \(V \d V^\dagger = -(\d V) V^\dagger \), obtained by differentiating \(V V^\dagger = \mathbbm{1}\).

        We further compute that
        \begin{equation}
            \tilde{n}_{V \tilde{U}} = -i(\tilde{U}^\dagger (V^\dagger\d V) \tilde{U} + \tilde{U}^\dagger \d \tilde{U})  = \tilde{U}^\dagger\tilde{n}_V \tilde{U} + \tilde{n}_{\tilde{U}} 
            \label{eqn:nVU}
        \end{equation}
        where
        \begin{equation}
            \tilde{n}_{\tilde{U}} = \sum_\alpha \Bn_\alpha \ketbra{\beta_\alpha}{\beta_\alpha}
        \end{equation}
        is a constant independent of \(\Bt\). Thus, substituting~\eqref{eqn:nVU} into the formula for the winding number~\eqref{eqn:wind2}, all the derivatives of \(\tilde{n}_{\tilde{U}}\) vanish, and we obtain
        \begin{equation}
            W[V\tilde{U}] = W[V] + B
        \end{equation}
        where
        \begin{align}
            B &= \tilde{C}_{d+D} \int \tr{\tilde{U} \tilde{n}_{\tilde{U}} \tilde{U}^\dagger \wedge (i\d\tilde{n}_V)^{\wedge (d+D-1)/2}} \\
            &= \tilde{C}_{d+D} \sum_\alpha \Bn_\alpha \cdot \bra{\alpha} \left(\int (i\d\tilde{n}_V)^{\wedge (d+D-1)/2}\right) \ket{\alpha}.
        \end{align}
        The integrand is a total derivative of
        \begin{equation}
            i \tilde{n}_V \wedge (i\d\tilde{n}_V)^{\wedge (d+D-3)/2},
        \end{equation}
        and so \(B = 0\).

        Thus, \(W[V\tilde{U}] = W[V]\) and the winding number is gauge invariant.

    \subsubsection{Proof of classification}
        \label{subsec:proof}

        We now prove the theorem classifying localized phases. We reproduce the theorem here.

        \begin{thm}
            The winding number \(W[V]\) is an integer valued topological invariant characterizing localized phases with \(d+D>1\). That is, if the two Hamiltonian-frequency pairs \((H_0(\Bt),\BO_0)\) and \((H_1(\Bt),\BO_1)\) are joined by a connected path \((H_s(\Bt),\BO_s)\) (where \(s\in [0,1]\)) such that all the \((H_s(\Bt),\BO_s)\) have localized quasienergy states, then \(W[V_0] = W[V_1]\).
        \end{thm}

        That \(W[V]\) is an integer and invariant under smooth deformations of \(V\) is well known, and we will assume this fact~\cite{Nakahara2003,Teo2010,Kitagawa2010,Yao2017}. More precisely, \(W[V]\) is a homotopy invariant. We show that, under the conditions of the theorem, the path between the micromotion operators \(V_s\) is continuous. Thus, the winding number of all micromotion operators on the path, in particular the end-points, must be equal.

        The proof is most straightforward in the frequency lattice. The continuous family \((H_s, \BO_s)\) defines a continuous family of quasienergy operators \(K_s\). The assumptions of the theorem require each \(K_s\) to have a complete set of normalizable eigenstates \(\ket{\tilde{\phi}_\gamma(s)}\) with associated eigenvalues \(\epsilon_\gamma(s)\), where \(\gamma\) indexes the frequency lattice Hilbert space. We show that eigenstate indices can be organized so that each \(\ket{\tilde{\phi}_\gamma(s)}\) is a continuous functions of \(s \in [0,1]\).

        Assuming for now that each \(\ket{\tilde{\phi}_\gamma(s)}\) is continuous in \(s\), the result follows straightforwardly as we have outlined. Any set of independent representatives for the quasienergy state equivalence classes (formed by~\eqref{eqn:quasi_gauge} and which we will indexed by \(\alpha\)) defines a continuous family of micromotion operators \(V_s = \sum_\alpha \ketbra{\phi_\alpha(s)}{\alpha}\). The homotopy invariance of \(W\) then implies that \(W[V_s] = W[V_{s'}]\) for all \(s\) and \(s'\). In particular \(W[V_0] = W[V_1]\).

        Organizing the eigenstates \(\ket{\tilde{\phi}_\gamma(s)}\) into continuous families is difficult (if it is possible) for general infinite dimensional gapless operators like \(K_s\). Localization allows us to construct these families with the same ease as in finite dimensional systems.

        We fix an eigenstate \(\ket{\tilde{\phi}_\gamma(s)}\) of \(K_s\), and consider eigenstates \(\ket{\tilde{\phi}_{\gamma'}(s')}\) of \(K_{s'}\) when \(s'\) is close to \(s\).
        
        First, observe that there are only finitely many \(\ket{\tilde{\phi}_{\gamma'}(s')}\) which can plausibly be matched to \(\ket{\tilde{\phi}_\gamma(s)}\). This is intuitive from localization: there are only finitely many quasienergy states localized near where \(\ket{\tilde{\phi}_\gamma(s)}\) has significant weight in the frequency lattice. Formally, we define a finite set of frequency lattice sites \(A\) such that
        \begin{equation}
            \bra{\tilde{\phi}_{\gamma}(s)} P_{A} \ket{\tilde{\phi}_{\gamma}(s)} > 1-\delta
        \end{equation}
        where \(P_A = \sum_{\Bn \in A} \ketbra{\Bn}{\Bn}\) is the projector onto \(A\) and \(\delta >0\). This is possible because \(\ket{\tilde{\phi}_{\gamma}(s)}\) is square-summable. This subset of the frequency lattice will be where we focus our attention.

        We consider a projection of \(K_{s'}\) onto \(A\), \(P_A K_{s'} P_A\). As \(K_{s'}\) is local, if \(\ket{\tilde{\phi}_{\gamma'}(s')}\) is almost entirely supported in the region \(A\), in the sense that \(\bra{\tilde{\phi}_{\gamma'}(s')} P_{A} \ket{\tilde{\phi}_{\gamma'}(s')} > 1-\delta\), then \(\ket{\tilde{\phi}_{\gamma'}(s')}\) is close to an eigenstate \(\ket{\chi_{\gamma'}(s')}\) of \(P_A K_{s'} P_A\). That is, if we write \(d(\psi,\phi)\) for the distance between states \(\ket{\psi}\) and \(\ket{\phi}\) (not being specific about the metric on state space), then for each \(\ket{\tilde{\phi}_{\gamma'}(s')}\) and any \(\epsilon >0\) there is a \(\delta>0\) (a large enough \(A\)) such that an eigenstate \(\ket{\chi_{\gamma'}(s')}\) of \(P_A K_{s'} P_A\) satisfies
        \begin{equation}
             d(\tilde{\phi}_{\gamma'}(s'), \chi_{\gamma'}(s')) < \epsilon.
             \label{eqn:chi_phi}
        \end{equation}

        Continuous families of eigenstates \(\ket{\chi_{\gamma'}(s')}\) can be unambiguously identified for the corresponding family of \emph{finite dimensional} hermitian operators \(P_A K_{s'} P_A\) (discarding the null space of \(P_A\)) for smooth enough paths~\cite[Chapter 2]{Kato1980}. This lets us make a choice for \(\gamma\) so that \(\ket{\chi_\gamma(s')}\) is continuous in \(s'\).

        The proximity of an eigenstate of \(K_{s'}\) to \(\ket{\chi_\gamma(s')}\), as in~\eqref{eqn:chi_phi}, then induces a choice for \(\ket{\tilde{\phi}_{\gamma}(s')}\). The family \(\ket{\tilde{\phi}_{\gamma}(s')}\) defined in this way is continuous: there is a \(\delta'>0\) so that \(d(\chi_{\gamma}(s), \chi_{\gamma}(s')) < \epsilon\) whenever \(|s-s'|<\delta'\) and so
        \begin{multline}
            d(\tilde{\phi}_{\gamma}(s),\tilde{\phi}_{\gamma}(s')) 
            \leq d(\tilde{\phi}_{\gamma}(s),\chi_{\gamma}(s)) + d(\chi_{\gamma}(s),\chi_{\gamma}(s')) \\
            + d(\chi_{\gamma}(s'),\tilde{\phi}_{\gamma}(s'))
            < 3 \epsilon,
        \end{multline}
        where we used the triangle inequality repeatedly and~\eqref{eqn:chi_phi}.

        This shows that the smooth path \((H_s, \BO_s)\) induces continuous paths for the frequency lattice eigenstates, and completes the proof of the theorem.

\subsection{Quantized Energy Circulation}
    \label{sec:quant_circ}

    We prove that the energy-charge circulation, generalizing the magnetization of the anomalous Floquet-Anderson insulator (AFAI)~\cite{Titum2016,Nathan2017}, is quantized and proportional to the winding number.

    For brevity of notation we will assemble the fluxes \(\Phi_j\) twisting the periodic boundary conditions of any spatial dimensions of the system and the drive phases \(\theta_{j}\) into a single three-dimensional vector \(\Bt\), as in~\eqref{eqn:phi_n_tht}. In this notation \(\BO\) is zero in any entry corresponding to a spatial dimension. Thus, we understand \(\Bt_t = \BO t + \Bt_0\) to only vary in time in the components corresponding to the drives.

    The component of the Heisenberg operator 
    \begin{equation}
        \dot{\Bn}(t;\Bt_0) = U^\dagger(t,0; \Bt_0) (-\nabla_{\theta} H)(\Bt_t) U(t,0; \Bt_0)
        \label{eqn:dotn_def}
    \end{equation}
    corresponding to drive \(j\) measures the photon current into drive \(j\). The component corresponding to the \(j\)th spatial axis measures a physical current. Thus, we identify its integral with a displacement in the frequency lattice,
    \begin{equation}
        \Delta\Bn(t;\Bt_0) = \int_0^t \d t\, \dot{\Bn}(t) = -i U^\dagger(t,0; \Bt_0) \nabla_{\theta_0} U(t,0; \Bt_0).
    \end{equation}
    This formula is most straightforwardly checked by differentiating the right hand side and applying the Schr\"odinger equation \(i\partial_t U = H U\) to obtain~\eqref{eqn:dotn_def}. Also observe that \(\Delta\Bn(0;\Bt_0) = 0\) as \(U(0,0;\Bt_0) = \mathbbm{1}\) is independent of initial phase.

    The component of \(\Delta\Bn\) corresponding to drive \(j\) is interpreted as the change in photon number of drive \(j\). The component corresponding to the \(j\)th spatial axis is interpreted as the displacement along this dimension, divided by the length of the system in that dimension, \(\Delta r_j/L_j\). An arbitrary choice of initial conditions for \(\Bn(t;\Bt_0) = \Bn(0;\Bt_0) + \Delta\Bn(t;\Bt_0)\) defines a position in the frequency lattice.

    The energy-charge circulation is defined by
    \begin{equation}
        M(t) = \frac{1}{4} (\Bn\times\dot{\Bn})\cdot\hat{\Omega} + \mathrm{h.c.}
        \label{eqn:quasi_mag}
    \end{equation}
    and we prove it takes the quantized average value
    \begin{equation}
        \ccexp{M}_T = \frac{1}{T}\int_0^T \d t\, \tr{M(t)} =  \frac{|\BO|}{2\pi} W[V] + O(T^{-1},e^{-L/\xi})
        \label{eqn:circ}
    \end{equation}
    in the anomalous localized phase for any initial phase \(\Bt_0\). Furthermore, we show that a local version of this quantity -- the circulation density -- is also quantized when one of the dimensions is spatial.

    As a preliminary issue, note that the long time average does not depend on the initial condition \(\Bn(0)\) in the localized phase. Adding an arbitrary constant to \(\Bn\) in~\eqref{eqn:quasi_mag} adds a term proportional to \(\frac{1}{T}\int_0^T \d t\, \dot{\Bn} = O(T^{-1})\), where we used that \(\Delta\Bn(t)\) is bounded in the localized phase.

    \subsubsection{Manipulation of winding number density}
        \label{subsec:wind_dens}

        Before we begin, it is convenient to first prove a lemma about the expression of the winding number density in terms of coordinates. The winding number density can be expressed in a coordinate-free form as
        \begin{equation}
            w[V] = \frac{i}{3! (2\pi)^2} \tr{\tilde{n} \wedge \tilde{n} \wedge \tilde{n} }
        \end{equation}
        where \(\tilde{n}(\Bt) = -i V^\dagger \d V\) and \(V(\Bt)\) is the micromotion operator~\eqref{eqn:micro}. To relate this to an expression with coordinates (such as \(M(t)\)) we prove
        \begin{equation}
            \tr{\tilde{n}_1 \partial_2 \tilde{n}_3}\, \d^3 \theta = -\frac{i}{6}\tr{ \tilde{n} \wedge \tilde{n} \wedge \tilde{n} } + \d \omega
            \label{eqn:wdens_coord}
        \end{equation}
        where subscript numerals index any local set of coordinates (which need not necessarily extend to a \emph{global} set of coordinates), we have defined a coordinate expression of \(\tilde{n}\) as \(\tilde{n} = \sum_{i=1}^3 \tilde{n}_i \d\theta_i\), \(\d \omega\) is a total derivative and \(\d^3\theta = \d \theta_1 \wedge \d\theta_2 \wedge \d\theta_3\) is the volume element of the torus. That is, we relate the coordinate-free expression on the right hand side of \eqref{eqn:wdens_coord} to a particular form involving the components of \(\tilde{n}\) on the left hand side.

        Indeed, we have
        \begin{align}
            \tilde{n}_1 \partial_2 \tilde{n}_3 &= -i \tilde{n}_1 \partial_2 (V^\dagger \partial_3 V) \\
            &= -i \tilde{n}_1 (\partial_2 V^\dagger \partial_3 V + V^\dagger \partial_2 \partial_3 V).
        \end{align}
        Inserting \(V^\dagger V = \mathbbm{1}\) in the first term and using \(V \partial V^\dagger = - \partial V V^\dagger\) (obtained by differentiating \(V V^\dagger = \mathbbm{1}\)) we see
        \begin{equation}
            \tilde{n}_1 \partial_2 \tilde{n}_3 = -i \tilde{n}_1 \tilde{n}_2 \tilde{n}_3 - i\tilde{n}_1 (V^\dagger \partial_2 \partial_3 V).
        \end{equation}
        The second term may be be further manipulated as
        \begin{multline}
            - i\tilde{n}_1 (V^\dagger \partial_2 \partial_3 V) = \partial_3 (\tilde{n}_1 \tilde{n}_2) - (\partial_3 \tilde{n}_1) \tilde{n}_2 + i \tilde{n}_1 (\partial_3 V^\dagger \partial_2 V) \\
            = \partial_3 (\tilde{n}_1 \tilde{n}_2) - (\partial_3 \tilde{n}_1) \tilde{n}_2 + i \tilde{n}_1 \tilde{n}_3 \tilde{n}_2.
        \end{multline}
        This gives the full expression
        \begin{equation}
            \tilde{n}_1 \partial_2 \tilde{n}_3 = \partial_3 (\tilde{n}_1 \tilde{n}_2) - (\partial_3 \tilde{n}_1) \tilde{n}_2 - i \tilde{n}_1 [\tilde{n}_2,\tilde{n}_3],
        \end{equation}
        which upon taking the trace and using the cyclic property thereof, becomes
        \begin{multline}
            \tr{\tilde{n}_1 \partial_2 \tilde{n}_3} = \partial_3 \tr{\tilde{n}_1 \tilde{n}_2} - \tr{\tilde{n}_2 \partial_3 \tilde{n}_1} \\
            - i \tr{\tilde{n}_1 [\tilde{n}_2,\tilde{n}_3]}.
        \end{multline}

        The first of these terms is a total derivative, and the final one appears in a coordinate expression of the winding number density. The second is of the same form as the left hand side (with the indices cyclically permuted), and so we may apply the same formula recursively. Doing this three times gives
        \begin{multline}
            \tr{\tilde{n}_1 \partial_2 \tilde{n}_3} = \partial_3 \tr{\tilde{n}_1 \tilde{n}_2} - \partial_1 \tr{\tilde{n}_2 \tilde{n}_3} + \partial_2 \tr{\tilde{n}_3 \tilde{n}_1}\\
             - \tr{\tilde{n}_1 \partial_2 \tilde{n}_3} - i \tr{\tilde{n}_1 [\tilde{n}_2,\tilde{n}_3]},
             \label{eqn:recurs3}
        \end{multline}
        where some terms have been canceled. The cyclicity of the trace may be further exploited to derive
        \begin{equation}
            \tr{\tilde{n}_1 [\tilde{n}_2,\tilde{n}_3]} = \frac{1}{3} \epsilon^{ijk} \tr{\tilde{n}_i \tilde{n}_j \tilde{n}_k},
            \label{eqn:com2levi}
        \end{equation}
        where summation over repeated indices is implied on the right hand side. The right hand side of~\eqref{eqn:com2levi} is proportional to the coefficient of \(\d^3\theta\) in \(\tr{ \tilde{n} \wedge \tilde{n} \wedge \tilde{n} }\). Moving the duplicate term \(-\tr{\tilde{n}_1 \partial_2 \tilde{n}_3}\) to the left hand side of~\eqref{eqn:recurs3} and multiplying by the volume element \(\d^3\theta\) gives the required expression
        \begin{equation}
            \tr{\tilde{n}_1 \partial_2 \tilde{n}_3}\, \d^3 \theta = -\frac{i}{6}\tr{ \tilde{n} \wedge \tilde{n} \wedge \tilde{n} } + \d \omega
        \end{equation}
        where
        \begin{multline}
            \omega = \frac{1}{2} \mathrm{Tr}\left[\tilde{n}_1 \tilde{n}_2 \,\d\theta_1\wedge\d\theta_2 \right.\\
            \left.- \tilde{n}_2 \tilde{n}_3 \,\d\theta_2\wedge\d\theta_3 + \tilde{n}_3 \tilde{n}_1 \,\d\theta_3\wedge\d\theta_1\right].
        \end{multline}

    \subsubsection{Proof of quantized energy-charge circulation}
        \label{subsec:circ_proof}

        We now use~\eqref{eqn:wdens_coord} to prove~\eqref{eqn:circ}. Portions of the following calculation are essentially a reproduction of the proof of quantized magnetization of the AFAI in~\cite{Nathan2017}.

        We define local coordinate vectors \(\hat{\ell}_1\) and \(\hat{\ell}_2\) which form an orthonormal triple with \(\hat{\Omega} = \hat{\ell}_1\times \hat{\ell}_2\). (Note that as \(\hat{\Omega}\) is incommensurate in general, these local coordinate vectors cannot be used to define a \emph{global} system of smooth coordinates on the torus.)

        First, we manipulate the formula for the winding number using~\eqref{eqn:wdens_coord}. The winding number density can be expressed in terms of our chosen coordinates as 
        \begin{equation}
            w[V] = \frac{-1}{(2\pi)^2} \tr{\tilde{n}_2  \partial_\Omega \tilde{n}_1 } \d^3\theta + \d\omega
            \label{eqn:wind_dens_coords}
        \end{equation}
        where we have denoted \(\partial_\Omega = \hat{\Omega} \cdot \nabla_{\theta}\). Integrating by parts and using the cyclicity of the trace,~\eqref{eqn:wind_dens_coords} becomes
        \begin{equation}
            w[V] = \frac{1}{8\pi^2} \tr{(\tilde{n}\times  \partial_\Omega \tilde{n})\cdot\hat{\Omega} } \d^3\theta +\d\omega'.
            \label{eqn:wind_dens}
        \end{equation}
        The winding number is the integral over the torus of this density, \(W[V] = \int_{\mathbb{T}^3} w[V]\), which removes the total derivative \(\d\omega'\).

        We replace the \(\Omega\) derivative by using the chain rule \(|\BO| \partial_\Omega A(\Bt_t) = \partial_t A(\Bt_t)\), valid for any \(A\) defined on the torus. We thus have for the winding number
        \begin{equation}
            |\BO| W[V] = \frac{1}{8\pi^2} \int\d^3\theta \,
            \tr{(\tilde{n}\times \dot{\tilde{n}})\cdot\hat{\Omega} }.
        \end{equation}

        The orbit \(\{\BO t + \Bt_0 \,:\, t \in \reals\}\) is ergodic in the torus when \(\BO\) is incommensurate. In the spatial dimensions where \(\BO\) has zero components, localization implies that the above quantity, which may be expressed as the trace of a Hermitian operator, depends only exponentially weakly on the threaded flux \(\Phi_j\). Thus, we may replace an average over all the variables \(\Bt\) with an average over just the orbit -- schematically \(\tfrac{1}{(2\pi)^3} \int \d^3 \theta = \tfrac{1}{T}\int_0^T \d t + O(T^{-1},e^{-L/\xi})\), where \(L\) is the linear system size and \(\xi\) is the localization length. For the winding number, we find
        \begin{multline}
            |\BO| W[V] = \frac{1}{8\pi^2} \frac{(2\pi)^3}{T}\int_0^T\d t \,
            \tr{(\tilde{n} \times \dot{\tilde{n}})\cdot\hat{\Omega}} \\
            + O(T^{-1},e^{-L/\xi}).
            \label{eqn:wind_tilde}
        \end{multline}

        We must now express~\eqref{eqn:wind_tilde} in terms of \(\Bn\), rather than \(\tilde{n}\). Using that \(V(\Bt_t) = U(t,0;\Bt_0) V(\Bt_0) e^{i H_F t}\) (as may be obtained from~\eqref{eqn:evo}) and that \(\d V(\Bt_t)\) has the components of \(\nabla_{\theta_0} V(\Bt_t)\), we have that 
        \begin{equation}
            \tilde{n}(\Bt_t) = e^{-iH_F t} V^\dagger_0 \Bn(t;\Bt_0) V_0 e^{i H_F t}.
            \label{eqn:tldn2n}
        \end{equation}
        Here we have denoted \(V(\Bt_0) = V_0\) and took for convenience that \(\Bn(0;\Bt_0) = V_0 \tilde{n}(\Bt_0) V^\dagger_0\). Taking the time derivative we find
        \begin{equation}
            \dot{\tilde{n}} = e^{-iH_F t} V^\dagger_0 \left( \dot{\Bn}(t) 
            + \com{\Bn(t)}{ V_0 H_F V^\dagger_0} \right) V_0 e^{i H_F t}.
            \label{eqn:tlddn2dn}
        \end{equation}

        Substituting~\eqref{eqn:tldn2n} and~\eqref{eqn:tlddn2dn} into~\eqref{eqn:wind_tilde} and focusing on the \(T \to \infty\) limit, we have
        \begin{align}
            \frac{|\BO|}{2\pi} W[V] = \lim_{T\to \infty} \frac{1}{T}\int_0^{T} \d t\, \left( \tr{\tfrac{1}{2}(\Bn \times \dot{\Bn}) \cdot \hat{\Omega}} \right. \label{eqn:line1}\\ 
            \left.+ \tr{\tfrac{1}{2}(\Bn \times i\com{\Bn}{V_0 H_F V_0^\dagger}) \cdot \hat{\Omega}} \right). \label{eqn:line3}
        \end{align}
        The first term~\eqref{eqn:line1} is the expression we are looking for.

        We must now argue that the last term~\eqref{eqn:line3} is zero. We observe that the remaining terms in~\eqref{eqn:line1} are invariant under the gauge transformation~\eqref{eqn:quasi_gauge}, so~\eqref{eqn:line3} must also be gauge invariant. We will show this is enough to conclude that it is, in fact, zero.

        We write \(V_0 H_F V_0^\dagger = \sum_{\alpha} \epsilon_\alpha \rho_\alpha\) where \(\rho_\alpha = \ketbra{\phi_\alpha(\Bt_0)}{\phi_\alpha(\Bt_0)}\). Then~\eqref{eqn:line3} becomes
        \begin{equation}
            \sum_\alpha \epsilon_\alpha \lim_{T\to \infty} \frac{1}{T}\int_0^{T} \d t\,\tr{\tfrac{1}{2}(\Bn \times i\com{\Bn}{\rho_\alpha}) \cdot \hat{\Omega}}.
            \label{eqn:line3b}
        \end{equation}
        Under a gauge transformation of \(\ket{\phi_\beta(\Bt)} \mapsto e^{i \Bk \cdot \Bt} \ket{\phi_\beta(\Bt)}\), the quasienergies transform as \(\epsilon_\alpha \mapsto \epsilon_\alpha + \Bk\cdot\BO \delta_{\alpha\beta}\), and the term~\eqref{eqn:line3b} is shifted by
        \begin{equation}
            \Bk\cdot\BO \lim_{T\to \infty} \frac{1}{T}\int_0^{T} \d t\,\tr{\tfrac{1}{2}(\Bn \times i\com{\Bn}{\rho_\beta}) \cdot \hat{\Omega}} = 0
            \label{eqn:gauge0}
        \end{equation}
        for any \(\Bk \in \integers^3\) and \(\beta\). Gauge invariance demands~\eqref{eqn:gauge0} is zero. As \(\Bk\cdot\BO \neq 0\) for at least some \(\Bk\), it is the second factor that must be zero for all \(\beta\). However, these are precisely the terms occurring in~\eqref{eqn:line3b}, so in fact~\eqref{eqn:line3b} is also zero.

        We are left with our desired term
        \begin{equation}
            \frac{|\BO|}{2\pi} W[V] = \lim_{T\to \infty} \frac{1}{T}\int_0^{T} \d t\, \tr{\tfrac{1}{2}(\Bn \times \dot{\Bn}) \cdot \hat{\Omega}}.
        \end{equation}
        The integrand is the trace of the product of two Hermitian operators, and so is real. Thus, we can take the hermitian part of the operator \(\tfrac{1}{2}(\Bn \times \dot{\Bn}) \cdot \hat{\Omega}\) in this formula. This is the energy-charge circulation~\eqref{eqn:quasi_mag}. Finally, we have
        \begin{equation}
            \frac{|\BO|}{2\pi} W[V] = \lim_{T\to \infty}\frac{1}{T}\int_0^{T} \d t\, \tr{M(t)}.
        \end{equation}

    \subsubsection{Quantized circulation density}
        \label{subsec:quant_dens}

        In addition to the total circulation~\eqref{eqn:quasi_mag} being quantized, the \emph{circulation density} is also quantized.

        The total circulation is defined using a trace over all states in a system with periodic boundary conditions. It should be regarded as the average of a locally defined circulation density for systems with at least one spatial dimension. In~\cite{Nathan2017} it was shown that not only is the magnetization quantized in a \((2+1)\)-dimensional system, but the magnetization \emph{density} is also quantized in a mesoscopic region filled with fermions. The equivalent claim in a \((1+2)\)-dimensional ALTP is also true; the circulation density is quantized, as we will now argue. The corresponding claim for a \((0+3)\)-dimensional ALTP is meaningless; with no spatial extent there is no sensible notion of density.

        \begin{figure}
            \centering
            \includegraphics{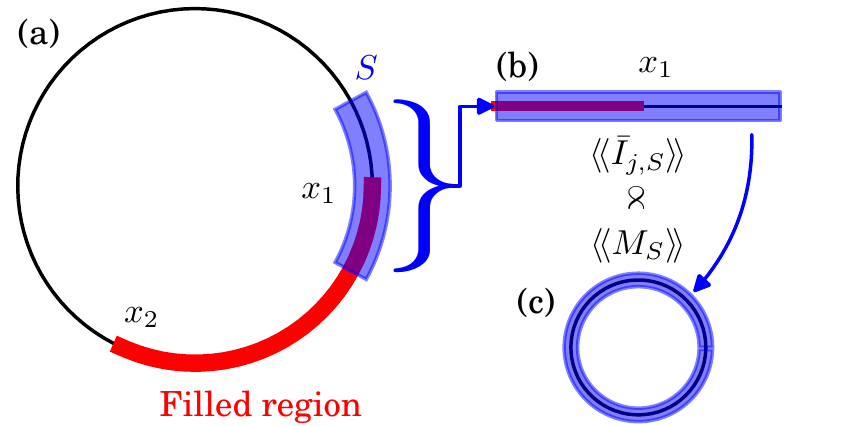}
            \caption{\label{fig:circdens}\emph{Geometry of \autoref{subsec:quant_dens}.---} \textbf{(a)} A large \((1+2)\)-dimensional ALTP, with all states in the red region filled with fermions. The energy current between the drives near one edge of the filled region (in blue), \(\ccexp{I_{j,S}}\), is equal (with exponentially small corrections) to the total energy current in a system with open boundary conditions obtained by cutting out the region \(S\) around \(x_1\), \(\ccexp{\bar{I}_{j,S}}\), shown in \textbf{(b)}. It is shown in \autoref{sec:quant_pump} that the current in \textbf{(b)} is proportional to the circulation \(\ccexp{M_S}\) in the system with periodic boundary conditions \textbf{(c)} obtained by joining the ends of \textbf{(b)} together. The total circulation of \textbf{(c)} must be equal to the circulation density in region \(S\) of \textbf{(a)}, as a localized system is insensitive to its boundary conditions. Thus the circulation of \textbf{(c)} is equal to the appropriately scaled average circulation density over \(S\), \(\tfrac{L}{|S|}\ccexp{M\rho_S}\). Then \(\ccexp{I_{j,S}} = -\ccexp{I_{j,S'}}\) (with \(S'\) near \(x_2\)) gives that \(\tfrac{L}{|S|}\ccexp{M\rho_{S}} = \tfrac{L}{|S'|}\ccexp{M\rho_{S'}}\).}
        \end{figure}

        We can calculate the average circulation density in a mesoscopic region by projecting the total circulation \(M\) onto some set of consecutive sites \(S\) with \(\xi \ll |S| \ll L\), where \(\xi\) is the localization length, \(|S|\) is the number of sites and \(L\) is the system size. That is, writing  \(m(r_1) = L \rho_{r_1} M \rho_{r_1}\) for the circulation density, where \(\rho_{r_1}\) is the projector into site \(\ket{r_1}\), and \(\ccexp{A} = \lim\limits_{T\to\infty} \int_0^T \d t\, \tr{A(t)}\) as usual, we have
        \begin{equation}
            \sum_{r_1 \in S} \ccexp{m(r_1)} = L \ccexp{M \rho_{S}},
        \end{equation}
        where \(\rho_S = \sum_{r_1 \in S} \rho_{r_1}\) is the projector onto the sites \(S\).

        We aim to show that 
        \begin{equation}
            \tfrac{L}{|S|}\ccexp{M \rho_S} = \tfrac{L}{|S'|}\ccexp{M \rho_{S'}} + O(e^{-|S|/\xi},e^{-|S'|/\xi})
        \end{equation}
        for any \(S\) and \(S'\) centered at \(x_1\) and \(x_2\) respectively. The result follows from a relation between the energy current between the drives and the circulation which we prove in \autoref{sec:quant_pump}, but will assume for now.

        Consider the average energy current into drive \(j\) in an initial state with fermions completely filling the region between \(x_1\) and \(x_2\):
        \begin{equation}
            \ccexp{I_j}_{[x_1,x_2]} = \lim_{T\to \infty}\frac{1}{T}\int_0^T \d t \, \tr{I_j(t) \rho_{[x_1,x_2]}}.
            \label{eqn:x1x2current}
        \end{equation}
        Here \(I_j(t) = \Omega_j \dot{n}_j(t)\) and \(\rho_{[x_1,x_2]}\) is the projector onto the sites between \(x_1\) and \(x_2\). Due to the localization of the quasienergy states, the only non-zero contributions to this integral can come from the ends of the filled region near \(x_1\) or \(x_2\). In the interior the localization of the quasienergy states ensures that \(\dot{\Bn}\) averages to zero, while the integrand is explicitly zero outside the filled region. In fact, the total integral must also be zero as the boundedness of \(n_j(t)\) in time in a localized phase implies that the average of \(\dot{n}_j(t)\) vanishes in any initial state.

        We can extract the energy current contribution to~\eqref{eqn:x1x2current} from the sites \(S\) near \(x_1\) by considering a system consisting only of the sites \(S\) with open boundary conditions (\autoref{fig:circdens}(a) and (b)). Due to the localization of the quasienergy states, the observable energy current in this segment only depends exponentially weakly on the different boundary conditions. Denoting operators on the system with open boundary conditions with a bar and writing \(\ccexp{I_{j,S}}\) for the energy current near \(x_1\) in the original system, we have
        \begin{equation}
            \ccexp{I_{j,S}} = \ccexp{\bar{I}_{j,S}} + O(e^{-|S|/\xi}).
        \end{equation}

        In \autoref{sec:quant_pump} we relate \(\ccexp{\bar{I}_{j,S}}\) to the average circulation of the system with periodic boundary conditions obtained by joining the ends of the open system (\autoref{fig:circdens}(b) and (c)). Calling this quantity \(\ccexp{M_S}\), we have 
        \begin{equation}
            \ccexp{\bar{I}_{j,S}} \propto \ccexp{M_S}.
        \end{equation}

        The circulation \(\ccexp{M_S}\) can now be related to our original quantity of interest \(\ccexp{M \rho_{S}}\). Indeed, localization implies the operators \(|S| M_S\) and \(L \rho_{S} M \rho_{S}\) coincide for states away from the boundaries of \(S\). The factors of system size \(|S|\) and \(L\) are present as the conjugate variable to the flux threading the small system (\autoref{fig:circdens}(c)) is \(r/|S|\), while in the large system (\autoref{fig:circdens}(a)) it is \(r/L\). Thus \(M\) carries an implicit factor \(1/L\), while \(M_S\) has a factor \(1/|S|\). Accounting for exponential corrections due to states localized near the boundary of \(S\), we have
        \begin{equation}
            \tfrac{L}{|S|}\ccexp{M \rho_{S}} = \ccexp{M_S} + O(e^{-|S|/\xi}).
        \end{equation}

        Following through the same logic at \(x_2\) with \(S'\), and paying careful attention to a minus sign due to the orientation at that boundary being opposite, we conclude that
        \begin{equation}
            \ccexp{I_j}_{s} \propto \tfrac{L}{|S|}\ccexp{M \rho_{S}} - \tfrac{L}{|S'|}\ccexp{M \rho_{S'}}+ O(e^{-s/\xi}).
        \end{equation}
        However, recall the left hand side must be zero as the average of \(\dot{n}_j(t)\) vanishes in any initial state. We then deduce that \(\tfrac{L}{|S|}\ccexp{M \rho_{S}} = \tfrac{L}{|S'|}\ccexp{M \rho_{S'}}\) up to exponentially small corrections. That is, the mesoscopic average of the circulation densities at \(x_1\) and \(x_2\) are equal.

\subsection{Quantized Edge Pumping}
    \label{sec:quant_pump}

    In addition to the quantized circulation of \autoref{sec:quant_circ}, \((1+2)\)-dimensional ALTPs also have topological edge effects. Namely, there is a quantized current of energy between the drives when the wire is prepared in an initial state with fermions filling all lattice sites near an edge (c.f.~\cite{Kolodrubetz2018,Peng2018b}).

\begin{figure}
        \centering
        \includegraphics{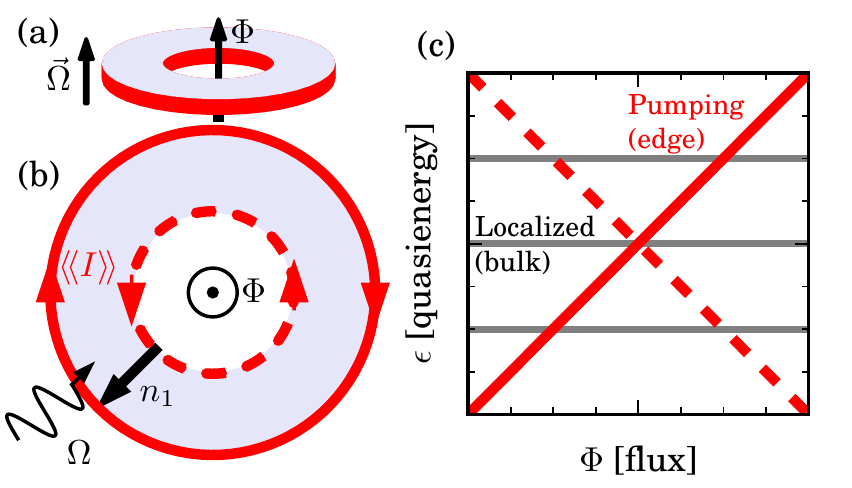}
        \caption{\label{fig:SpecFlow}\emph{Edge states in a commensurate approximation.---} \textbf{(a)} The \((3+0)\)-dimensional frequency lattice corresponding to a one-dimensional system driven by two \emph{commensurate} tones is a cylinder. \textbf{(b)} The corresponding \((2+1)\)-dimensional system is a driven annulus. A charge current \(\ccexp{I}\) in the \((2+1)\)-dimensional system corresponds to an energy current between the drives \(\ccexp{I_j}\) in the \((1+2)\)-dimensional system. \textbf{(c)} The spectral flow of the quasienergy states as a quantum of flux twists the periodic boundary conditions. The edge states responsible for the energy current wind around the domain of periodicity of the quasienergy, while the localized bulk states are unaffected.}
    \end{figure}

    The presence and nature of the edge states can be deduced intuitively by considering a commensurate approximation to the incommensurately driven problem of interest. As noted in \autoref{subsec:freq_lat_multi}, this commensurate approximation compactifies the frequency lattice model into a cylinder, which may be threaded by a flux \(\Phi\) (\autoref{fig:SpecFlow}(a)). The \((2+1)\)-dimensional ALTP (the AFAI) of equivalent geometry consists of a driven annulus and possesses edge states which carry a charge current along the two rings of the annulus (\autoref{fig:SpecFlow}(b)). The movement of charge around this cylinder in the AFAI corresponds in the frequency lattice to the transport of a state through different photon occupation states \(\ket{\Bn}\). That is, a current of energy between the drives.

    The same conclusion may be drawn by inspecting the quasienergy band structure of the \((2+1)\)-dimensional model. The quasienergies of the edge states wind \(W[V]\) times around their domain of periodicity as a quantum of flux twists the periodic boundary conditions. In the quasiperiodic limit (the infinite system size limit of the periodic dimension) the dependence of the gradient of the quasienergy on the threaded flux disappears~\cite{Crowley2019}. This results in quasienergy bands of constant gradient proportional to \(W[V]\) (\autoref{fig:SpecFlow}(c)). The gradient of the quasienergy itself is directly proportional to the long-time average of the pumped power~\cite{Crowley2019}.

    With this intuition we now proceed to the formal proof, which does not make use of any commensurate approximation. The proof, not surprisingly, mirrors the corresponding proof for the AFAI~\cite{Nathan2017}.

    We consider a one-dimensional lattice of sites \(\ket{n_1}\) driven by two incommensurate tones \(\BO = \Omega_2 \hat{e}_2 + \Omega_3 \hat{e}_3\). The winding number \(W[V]\) invariant of this system is associated with periodic boundary conditions in the \(n_1\) direction, being given explicitly by~\eqref{eqn:wind_num}. We will show the winding number \(W[V]\) is related to energy current at the edge of a chain with \emph{open} boundary conditions, but which is identical to the periodic system in the bulk.

    Unlike \autoref{sec:quant_circ}, we will use the standard coordinate axes, so \(\partial_1 = \partial_{\Phi_1}\) (with \(\Phi_1\) being a flux) while \(\partial_{2,3} = \partial_{\theta_{02,03}}\). As in \autoref{sec:quant_circ}, we write \(n_j(t;\Bt_0) = n_j(0;\Bt_0) -i U^\dagger \partial_j U(t,0;\Bt_0) \) for the frequency lattice position along the \(j\) axis, and \(\dot{n}_j(t;\Bt_0) = -\bar{U}^\dagger \partial_{j}\bar{H}(\Bt_t) \bar{U}\) for its derivative. We will denote operators in the system with open boundary conditions by a bar, so that the open system has Hamiltonian \(\bar{H}\), micromotion \(\bar{V}\) and so on.

    We will prove that the average energy current into drive two (three) is quantized when the lattice is initialized in a state with fermions completely filling all sites localized near one of the edges. That is,
    \begin{multline}
        \ccexp{\bar{I}_{2,3}}_{s,T} \equiv \frac{1}{T}\int_0^T \d t \, \tr{\bar{I}_{2,3}(t) \rho_s} \\
        = \pm\frac{\Omega_2 \Omega_3}{2\pi} W[V] + O(T^{-1},e^{-s/\xi},e^{-L/\xi})
        \label{eqn:power}
    \end{multline}
    where \(\bar{I}_j(t) = \Omega_j \dot{\bar{n}}_j(t)\) (\(j \in \{2,3\}\)) is the Heisenberg operator for the current into drive \(j\), \(\rho_s\) is a projector onto the sites \(\ket{n_1}\) within a distance \(s\) of the edge, \(L\) is the length of the chain and \(\xi\) is the single-particle localization length.

    First we relate \(\bar{I}_j(t)\) to a Berry curvature. This calculation is standard in the literature~\cite{Rudner2013,Titum2016,Martin2017,Crowley2019}. Using the product rule, we have
    \begin{equation}
        -\bar{U}^\dagger \partial_j \bar{H} \bar{U} = -\partial_j(\bar{U}^\dagger \bar{H}\bar{U}) + (\partial_j\bar{U}^\dagger) \bar{H}\bar{U} + \bar{U}^\dagger \bar{H} (\partial_j\bar{U}).
    \end{equation}
    Using the Schr\"odinger equation \(i\partial_t \bar{U} = \bar{H} \bar{U}\), this becomes
    \begin{equation}
        -\bar{U}^\dagger \partial_j \bar{H} \bar{U} = -i\partial_j(\bar{U}^\dagger \partial_t\bar{U}) + i(\partial_j\bar{U}^\dagger \partial_t\bar{U} - \partial_t\bar{U}^\dagger \partial_j\bar{U}).
        \label{eqn:berry_curv}
    \end{equation}
    
    The first term, when substituted back into the integral gives a contribution proportional to \(\partial_j \left(\tfrac{1}{T}\int\d t\, \tr{\bar{U}^\dagger \bar{H}\bar{U}\rho_s}\right)\), which is the \(\theta_{0j}\) derivative of the average energy. In the \(T\to \infty\) limit the average energy becomes insensitive to the initial phase, and so this term is zero. This can be seen by first noting that in the bulk of the filled region the integral becomes \(\tfrac{1}{T}\int\d t\, \partial_j\tr{\bar{H}} = \tfrac{1}{(2\pi)^3} \int \d^3\theta\, \partial_j\tr{\bar{H}} + O(T^{-1})\) which is the integral of a total derivative over the torus, and so is zero. Away from the boundary of the lattice all the quasienergy states are localized in the frequency lattice, and so their instantaneous energy also has the periodicity of the torus, and their derivatives integrate to zero.

    The remaining term in~\eqref{eqn:berry_curv} is indeed a Berry curvature. By inserting \(\bar{U} \bar{U}^\dagger = \mathbbm{1}\) and using \((\partial \bar{U}^\dagger) \bar{U} = - \bar{U}^\dagger \partial\bar{U}\) we put the full expression~\eqref{eqn:power} in the form
    \begin{equation}
        \ccexp{\bar{I}_j}_{s,T} = -\frac{i \Omega_j}{T}\int_0^T \d t \, \tr{\com{\bar{n}_t}{\bar{n}_j} \rho_s}
    \end{equation}
    where we denoted \(\bar{n}_k = -i \bar{U}^\dagger \partial_k \bar{U}\). Using the fact that \(\tr{\com{A}{B} C} = -\tr{B\com{A}{C}}\) this is
    \begin{equation}
        \ccexp{\bar{I}_j}_{s,T} = \frac{i \Omega_j}{T}\int_0^T \d t \, \tr{\bar{n}_j [\bar{n}_t,\rho_s]}.
        \label{eqn:P_with_com}
    \end{equation}

    \autoref{eqn:P_with_com} may be related to the model with periodic boundary conditions through the use of an auxiliary gauge transformation of the form
    \begin{align}
        \ket{n_1} \mapsto \ket{n_1},\quad &n_1 \leq s \\
        \ket{n_1} \mapsto e^{i\Phi_1}\ket{n_1},\quad &n_1 > s
    \end{align}
    which is implemented by the unitary \(G_{\Phi_1} = e^{i \Phi_1 (\mathbbm{1}-\rho_s)}\). As this is a pure gauge transformation on the system with open boundary conditions in the \(n_1\) direction, it does not affect expectation values of physical observables. Then, defining \(\bar{A}(\Phi_1) = G_{\Phi_1}^\dagger \bar{A} G_{\Phi_1}\) and using that \(G_{\Phi_1}\) and \(\rho_s\) commute we have
    \begin{equation}
        \ccexp{\bar{I}_j}_{s,T} = \frac{i \Omega_j}{T}\int_0^T \d t \,
        \tr{\bar{n}_j(\Phi_1) [\bar{n}_t(\Phi_1),\rho_s]}.
    \end{equation}
    The commutator in this expression can be expressed as a derivative,
    \begin{equation}
        [\bar{n}_t(\Phi_1),\rho_s] = i\partial_1 \bar{n}_t(\Phi_1),
    \end{equation}
    giving the expression
    \begin{equation}
        \ccexp{\bar{I}_j}_{s,T} = -\frac{\Omega_j}{T}\int_0^T \d t \,
        \tr{\bar{n}_j(\Phi_1) \partial_1 \bar{n}_t(\Phi_1)}.
        \label{eqn:chain_pow}
    \end{equation}

    We now relate this expression involving operators on the chain with open boundary conditions (with bars) to a similar expression on the system with periodic boundary conditions (with no decorations). That is, we consider a one-dimensional system with periodic boundary conditions, a circle, driven by two periodic tones with a flux \(\Phi_1\) twisting the boundary conditions. The Hamiltonian \(H(t;\Phi_1,\theta_{02},\theta_{03})\) on the circle is identical to \(G_{\Phi_1}^\dagger \bar{H}(t;\theta_{02},\theta_{03})G_{\Phi_1}\) in the interior of the chain.

    Due to the assumed localization of the bulk of the chain, the operators \(\bar{n}_k\) are all themselves local. This means that on the circle \(n_t(\Phi_1)\) depends only exponentially weakly on \(\Phi_1\). This weak dependence on the twisted boundary condition \(\Phi_1\) and the posited matching of \(H\) and \(\bar{H}\) in the interior implies \(i\partial_1 \bar{n}_t(\Phi_1) = i\partial_1 n_t(\Phi_1) + O(e^{-L/\xi})\). We also see from~\eqref{eqn:P_with_com} that the only states making significant contribution to the integrand are those near \(n_1 = s\) -- this is the only region where \(\rho_s\) is not locally proportional to the identity, and so is the only place the commutator can be non-zero. For large enough \(s\) this is well within the bulk of the chain, where the \(\bar{n}_k\) operators match the operators on the circle. We conclude that we may replace the chain operators in~\eqref{eqn:chain_pow} with their periodic boundary condition equivalents, with only exponentially small corrections in \(L/\xi\) and \(s/\xi\), which we suppress. This gives
    \begin{equation}
        \ccexp{\bar{I}_j}_{s,T} = -\frac{\Omega_j}{T}\int_0^T \d t\, \tr{n_j(\Phi_1) \partial_1 n_t(\Phi_1)}.
        \label{eqn:ring_pow}
    \end{equation}

    The integrand here is an expression to which the result of \autoref{subsec:wind_dens}~\eqref{eqn:wdens_coord} applies. This result may be applied repeatedly to show (suppressing \(\Phi_1\) dependence)
    \begin{equation}
        \tr{n_j\partial_1 n_t} = -\tr{n_1\partial_t n_j} + \partial_j \tr{n_1 n_t}.
    \end{equation}
    The product rule then gives
    \begin{equation}
        \tr{n_1\partial_t n_j} = \tfrac{1}{2}\tr{n_1\partial_t n_j - n_j\partial_t n_1} + \tfrac{1}{2}\partial_t \tr{n_j n_1},
    \end{equation}
    so that the integrand is
    \begin{multline}
        -\tr{n_j\partial_1 n_t} = \tfrac{1}{2}\tr{n_1\partial_t n_j - n_j\partial_t n_1} \\
        - \partial_j \tr{n_1 n_t} + \tfrac{1}{2}\partial_t \tr{n_j n_1}.
        \label{eqn:curr_cross}
    \end{multline}
    The first of these terms is \(\pm \tr{\tfrac{1}{2}(\Bn \times \dot{\Bn})\cdot\hat{e}_k}\), where \(k \neq j\) and the sign is positive for \(j=2\) and negative for \(j = 3\).

    Substituting~\eqref{eqn:curr_cross} into~\eqref{eqn:ring_pow}, we see
    \begin{multline}
        \ccexp{\bar{I}_j}_{s,T} = \pm\frac{\Omega_j}{ T}\int_0^T \d t\, 
        \tr{\tfrac{1}{2}(\Bn \times \dot{\Bn})\cdot\hat{e}_k} \\
        - \partial_j \tr{n_1 n_t} + \partial_t \tr{\tfrac{1}{2}n_j n_1}.
        \label{eqn:tot_deriv}
    \end{multline}

    Both of the total derivative terms average to zero in the \(T \to \infty\) limit. This can be seen explicitly for the time derivative:
    \begin{equation}
        \frac{1}{T}\int_0^T \d t\, \partial_t \tr{\tfrac{1}{2} n_j n_1} = \frac{1}{T} \left.\tr{\tfrac{1}{2} n_j n_1}\right|_0^T = O(T^{-1})
    \end{equation}
    as \(\Bn\) is bounded in time by the assumed localization in both space and the frequency lattice.

    The other derivative term is an initial phase derivative of the long-time average of an observable, and so is also zero. Explicitly, expanding the trace in a basis of quasienergy states:
    \begin{multline}
        \tr{n_1 n_t} = \sum_{\alpha,\beta} \bra{\psi_\alpha(t)} \partial_1 \ket{\psi_\beta(t)} \bra{\psi_\beta(t)} \partial_t \ket{\psi_\alpha(t)} \\
        = \sum_{\alpha,\beta} \left(\bra{\phi_\alpha(\Bt_t)} \partial_1 \ket{\phi_\beta(\Bt_t)} \bra{\phi_\beta(\Bt_t)} \partial_t \ket{\phi_\alpha(\Bt_t)}\right. \\
        \left.- i\epsilon_\alpha \bra{\phi_\alpha(\Bt_t)} \partial_1 \ket{\phi_\alpha(\Bt_t)} \delta_{\alpha\beta} \right),
    \end{multline}
    which is periodic on the torus \(\Bt_t\), and so its initial phase derivatives vanish upon averaging.

    We are left with 
    \begin{equation}
        \ccexp{\bar{I}_j}_{s,T} = \pm \Omega_j \ccexp{\tfrac{1}{2}(\Bn \times \dot{\Bn})\cdot\hat{e}_k}_T.
        \label{eqn:Ij_ek}
    \end{equation}
    We want to relate this to \(\ccexp{\tfrac{1}{2}(\Bn \times \dot{\Bn})\cdot\hat{\Omega}}_T = \ccexp{M}_T\), so that we can use the main result of \autoref{sec:quant_circ}~\eqref{eqn:circ} to relate this to the winding number. The spatial system is localized, and so cannot absorb energy indefinitely. Thus, the long-time average of the energy current into the system is zero. That is,
    \begin{multline}
        \ccexp{\tfrac{1}{2}(\Bn \times \dot{\Bn})\cdot(-\Omega_2\hat{e}_3 +\Omega_3\hat{e}_2)}_T
        = -\ccexp{\bar{I}_2}_{s,T} - \ccexp{\bar{I}_3}_{s,T} \\
        =\ccexp{\partial_t\bar{H}}_{s,T} = O(T^{-1}).
    \end{multline}
    Writing \(\hat{\ell} = (-\Omega_2\hat{e}_3 +\Omega_3\hat{e}_2)/|\BO|\), we see that \(\ccexp{\tfrac{1}{2}(\Bn \times \dot{\Bn})}_T\) has no component along \(\hat{\ell}\) in the long time limit. Then we can decompose \(\hat{e}_k = (\hat{\ell}\cdot\hat{e}_k) \hat{\ell} + (\hat{\Omega}\cdot\hat{e}_k) \hat{\Omega}\) in~\eqref{eqn:Ij_ek}, and discard the part with \(\hat{\ell}\). This leaves just
    \begin{equation}
        \ccexp{\bar{I}_j}_{s,T} = \pm \frac{\Omega_j \Omega_k}{|\BO|}  \ccexp{\tfrac{1}{2}(\Bn \times \dot{\Bn})\cdot\hat{\Omega}}_T + O(T^{-1}).
        \label{eqn:Ij_omg}
    \end{equation}

    Using the main result of \autoref{sec:quant_circ}, this is
    \begin{equation}
        \ccexp{\bar{I}_j}_{s,T} = \pm \frac{\Omega_2 \Omega_3}{2\pi} W[V] + O(T^{-1},e^{-s/\xi},e^{-L/\xi}),
    \end{equation}
    where we restored the suppressed exponential corrections.

\subsection{Description of numerics}
    \label{sec:numerics}

    In this section we describe the numerical experiments reported in the main text in detail.

    Time evolution in both zero- and one-dimensional driven models was implemented using the second-order of the Suzuki-Trotter decomposition of the unitary evolution operator \(U(t_1, t_0; \Bt_0)\)~\cite{Wiebe2010}. That is, defining
    \begin{equation}
        U_1(t_1,t_0; \Bt_0) = \exp(-i\Delta t H(\Bt_{(t_1+t_0)/2}))
    \end{equation}
    as a first order approximation to the evolution, with \(\Delta t = t_1-t_0\), and \(s_k = (4-4^{1/(2k+1)})^{-1}\), we approximated
    \begin{multline}
        U_2(t_1, t_0;\Bt_0) = U_1(t_1, t_1-s_1\Delta t)
        U_1(t_1- s_1\Delta t, t_1- 2s_1\Delta t) \\
        U_1(t_1- 2s_1\Delta t, t_0+2s_1\Delta t)
        U_1(t_0+2s_1\Delta t, t_0+s_1\Delta t) \\
        U_1(t_0+s_1\Delta t, t_0),
    \end{multline}
    where we have suppressed \(\Bt_0\) dependence on the right hand side. This method explicitly ensures the unitarity of evolution. Furthermore, with our choice of \(U_1\) it accumulates error due to the variation of \(H(\Bt_t)\) as a function of \(\Bt_t\), rather than the variation of the state \(\ket{\psi(t)}\) as a function of time. Indeed, for a static Hamiltonian this method is exact to within numerical precision if \(U_1\) can be exactly calculated.

    For the qubit model \(H^\delta(\Bt_t)\), we calculated the full \(2\times2\) approximation to the evolution operator
    \begin{equation}
        U(t_n,0;\Bt_0) \approx \prod_{j=1}^n U_2(t_j,t_{j-1};\Bt_0)
    \end{equation}
    at times \(t_n = n \Delta t\) up to a maximum of \(T\), where \(n\) is an integer and \(\Delta t = 2\pi/(50 \Omega_1)\) (which we found to be sufficient for convergence on numerical timescales).

    To calculate \(\ccexp{M}_T\) we compute the Heisenberg operator \(\dot{\Bn}(t_n) = U^\dagger (-\nabla_{\Bt} H^\delta) U(t_n)\) and integrate it with the trapezoidal rule to find \(\Bn(t_n)\). We take initial conditions \(\Bn(0) = 0\). As observed in \autoref{sec:quant_circ}, this does not affect the long-time value of \(\ccexp{M}_T\) in a localized phase. With \(\dot{\Bn}\) and \(\Bn\), it is straightforward to compute \(\ccexp{M}_T\) by taking the trace of the triple product of \(\Bn\), \(\dot{\Bn}\) and \(\hat{\Omega}\)~\eqref{eqn:quasi_mag}.

    We non-dimensionalize calculations with \(H^\delta(\Bt_t)\) by fixing units of energy so that \(B_0 = 1\). In Fig.~2(a,b) we use the parameters \(\delta = 0.01\), \(\vec{\omega} = (2, 1.618031..., 1.073506...)\) and \(\BO \propto \vec{\omega}\) for \(H^\delta\).

    The phase diagram Fig.~2(a) shows the value of \(\ccexp{M}_T\) (as colors) as computed above for \(T = 2^{14} (2\pi/\Omega_1)\) for a range of \(h\) and \(|\BO|/|\vec{\omega}|\). \(\ccexp{M}_T\) was calculated for \(n_\theta = 16\) random initial phases \(\Bt_0\). In the localized phase, these should all converge to the same quantized value. If the samples failed to converge on our numerical timescales, the corresponding point in the phase diagram was colored black. Away from transitions, the calculated \(\ccexp{M}_T\) is quantized. Closer to transitions, smaller step sizes \(\Delta t\) and longer integration times \(T\) are required to see precise quantization, as finite-time effects become more significant.

    The time series of \(\ccexp{M}_T\) in Fig.~2(b) were calculated with the same parameters as Fig.~2(a), but at specific points with \(\BO = \vec{\omega}\) and \(h \in \{2,5\}\). The plots show the average of \(\ccexp{M}_T\) over \(n_\theta = 64\) random initial phases, with error bars (usually too small to see) showing the standard deviation. The asymptotes of both plots are quantized at the predicted values.

    For numerics involving a two-tone driven wire (Figs.~2(c) and 3), we used sparse matrix methods to compute \(U_2(t_n,t_{n-1};\Bt_0)\ket{\psi(t_{n-1})}\) from initial states \(\ket{\psi(0)}\) localized near an edge. Our model in this case is obtained from \(H^\delta(\Bt)\) as used in Fig.~2(b) via Fourier transform, as described in the main text.

    The time series of Fig.~2(c) show the total work done \(T\ccexp{I_j}_{s,T}\) on the drives for the two parameter values of Fig.~2(b). The lattice size is \(L = 40\), and the \(s=14\) sites closest to the edge are initially filled with fermions. Faint lines are plotted showing \(T\ccexp{I_j}_{s,T}\) for each of \(n_\theta = 8\) initial phases, and a thick line shows the average. The faint trajectories all remain close together (frequently obscured by the thick trajectory), consistent with our prediction that \(T\ccexp{I_j}_{s,T} \sim \tfrac{\Omega_2 \Omega_3}{2\pi} W[V] T\) regardless of initial phase. This prediction is also plotted in Fig.~2(c).

    The predicted quantization of \(\ccexp{I_j}_{s,T}\) is shown in Fig.~3, as is the exponential localization of the edge modes. Each data point in Fig.~3 represents a linear fit to the work done \(T\ccexp{I_j}_{s,T}\) for a different \(s\), using the same model and parameters as Fig.~2(c). We call the fitted slope \(\ccexp{I_j}_{s,\mathrm{fit}}\), and see that it saturates to the predicted value in both the \(W[V] = 1\) and \(W[V] = 0\) phase. An exponential fit to \(\ccexp{I_j}_{s,\mathrm{fit}}\) in the \(W[V] = 1\) phase gives a localization length \(\xi \approx 2\).

\subsection{Qudit-cavity ALTP}
    \label{sec:qcav}

    In the main text our primary model for a \((1+2)\)-dimensional ALTP was of a quasiperiodically driven wire of non-interacting fermions. The single-particle Hilbert space for a spinful fermion hopping on a semi-infinite one-dimensional lattice is identical to the Hilbert space of a spin coupled to a quantum cavity, where now the vacuum state functions as an edge. As the responses we have discussed -- quantized circulation and energy currents -- are all controlled by single-particle physics, we can observe the same effects in a qubit coupled to a cavity.

    In fact, a cavity model is naturally constructed from the frequency lattice of any \((0+3)\)-dimensional ALTP. The frequency lattice can be considered to be the high-photon limit of \(D=3\) quantum cavities coupled to a qudit. In this picture the ``electric potential'' \(\BO\cdot\Bn\) is reinterpreted as the sum of cavity energies \(\Omega_j n_j\) (where \(\hbar = 1\)). Then, when seeking a \((1+2)\)-dimensional model with an edge, it is intuitive to take this limiting construction of the frequency lattice and replace one synthetic dimension in the frequency lattice with a cavity, where Fock states are represented by lattice sites.

    Explicitly, given a Hamiltonian \(H(\Bt) = \sum_{\Bn} H_{\Bn} e^{-i\Bn\cdot\Bt}\) for a \((0+3)\)-dimensional ALTP, we construct a \((1+2)\)-dimensional cavity model by quantizing one of the classical drives (without loss of generality, drive 1). That is, we make the replacement \(e^{i \theta_1} \mapsto a/\sqrt{n_0}\) in \(H(\Bt)\), where \(a\) is the cavity annihilation operator and \(n_0\) fixes an energy scale. Then the qudit-cavity Hamiltonian is
    \begin{multline}
        H_c(\theta_2,\theta_3) = \sum_{\{\Bn\,:\,n_1>0\}} H_{\Bn} \left(\frac{a^{\dagger}}{\sqrt{n_0}}\right)^{|n_1|} e^{-i(n_2\theta_2 + n_3\theta_3)} \\
        + \sum_{\{\Bn\,:\,n_1<0\}} H_{\Bn} \left(\frac{a}{\sqrt{n_0}}\right)^{|n_1|} e^{-i(n_2\theta_2 + n_3\theta_3)}.
    \end{multline}

    As claimed, this model coincides with the frequency lattice model in the particular high photon number limit \(m_1 \to \infty\) with \(m_1/n_0 \to 1\), where \(\ket{m_1}\) is a Fock state of the cavity. Away from this strict limit, the cavity model differs from the frequency lattice model due to the Bose enhancements \(\sqrt{m_1/n_0}\) on the creation and annihilation operators \(a^{(\dagger)}/\sqrt{n_0}\). These Bose enhancement factors are the origin of some subtlety, which we now expand upon.

    In the vicinity of the Fock state \(\ket{m_1}\) the Hamiltonian has \(n\)-site hopping matrices approximately given by \(H_{\Bn} (m_1/n_0)^{|n_1|/2}\). Clearly, these matrices do not coincide with those in the original frequency lattice model except in the vicinity of \(m_1 = n_0\). We make a kind of local density approximation (LDA) by treating quantities of interest (in particular the winding number) as functions of \(m_1\), and calculating them as if in an infinite, uniform model with hops given by \(H_{\Bn} (m_1/n_0)^{|n_1|/2}\). This LDA model, if it is still localized, has an associated winding number \(W(m_1)\).

    \begin{figure}[b]
        \centering
        \includegraphics{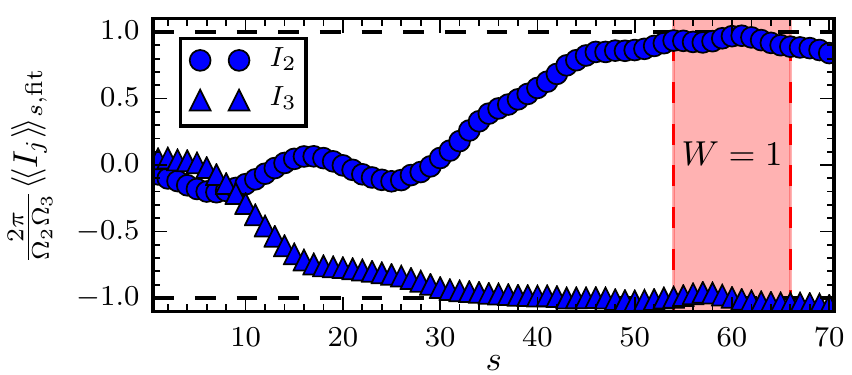}
        \caption{\label{fig:edge_cav}\emph{Qubit-cavity energy current.---} The slope \(\ccexp{I_j}_{s,\mathrm{fit}}\) is fitted to the sum of the work done on each drive in initial Fock states \(\ket{m_1}\) with \(m_1 < s\). The model used is obtained from \(H^\delta\) of the main text by making the replacement \(e^{i \theta_1} \mapsto a/\sqrt{n_0}\) where \(a\) is the cavity annihilation operator. A uniform model characterizing the local behavior of the cavity model may be constructed by taking the local hopping matrices \(H_{\Bn} (m_1/n_0)^{|n_1|/2}\) of the cavity model and repeating them, forming a local density approximation (LDA). The estimated region where this model produces an ALTP of winding number \(W = 1\) is shaded in red in the figure. Within the core of this region the energy current is quantized at the predicted value. \emph{Parameters:} \(\delta/B_0 = 0.01\), \(\BO=\vec{\omega} = B_0(2, 1.618031..., 1.073506...)\), \(h=2\), \(n_0 = 60\). The cavity Fock space was truncated at \(m_{\max} = 100\).}
    \end{figure}

    It is unlikely that \(W(m_1)\) is constant in the entire range \(0 \leq m_1 \leq n_0\) if \(W(n_0)\) is nontrivial. Usually, the suppression of the hopping matrices by the factor \((m_1/n_0)^{|n_1|/2}\) results in either a different winding number, or the delocalization of the quasienergy states. When seeking to observe topological edge effects, the entire region between distinct winding numbers should be regarded as composing the ``edge'' between topological regimes. The states responsible for the quantized energy current of \autoref{sec:quant_pump} may be delocalized in this entire region.

    \autoref{fig:edge_cav} shows the energy current in states supported within \(s\) sites of the vacuum state \(\ket{0}\) for a particular qubit-cavity ALTP, similar to Fig.~3 of the main text. The cavity model used in the figure was obtained by quantizing the first drive of the qubit model \(H^\delta(\Bt)\) in the main text with \(n_0 = 60\). The total current is nearly quantized in the topological regime where \(W(m_1) = 1\), which is marked in \autoref{fig:edge_cav}. The lack of quantization of the current outside this region indicates that the LDA model's quasienergy states are delocalized. Without localization the current may in principle take any value.

    Depending on the experimental architecture in question, constructing \((1+2)\)-dimensional ALTPs using a cavity coupled to a qubit may be easier than accomplishing the same task in a wire of fermions~\cite{Ozawa2019a}. In our particular model, \(H^\delta\), the topological region in Fock space tends to be quite narrow, which is not desirable for an experimental realization. It would be useful to find new models with a more robust range of non-trivial \(W(m_1)\).

\end{document}